\documentclass[pre,twocolumn]{revtex4-1}
\usepackage{graphicx}
\usepackage{epstopdf}
\usepackage{amsmath}
\usepackage{amssymb}
\usepackage{booktabs}
\usepackage{color}

\begin{document}

\title{Purely hydrodynamic ordering of rotating disks \\
at a finite Reynolds number} 

\author{Yusuke Goto}
\author{Hajime Tanaka \footnote{Correspondence and requests for materials should be addressed to 
H. T. (tanaka@iis.u-tokyo.ac.jp).}}
\affiliation{ {Institute of Industrial Science, University of Tokyo, 4-6-1 Komaba, Meguro-ku, Tokyo 153-8505, Japan} }

\date{Received August 14, 2014}

\begin{abstract}
{\bf
Self-organization of moving objects in hydrodynamic environments has recently attracted considerable attention in connection to natural phenomena and living systems. However, the
underlying physical mechanism is much less clear due to the intrinsically nonequilibrium nature, compared with self-organization of thermal systems. Hydrodynamic interactions are
believed to play a crucial role in such phenomena. To elucidate the fundamental physical nature of many-body hydrodynamic interactions at a finite Reynolds number, here we study a
system of co-rotating hard disks in a two-dimensional viscous fluid at zero temperature. Despite the absence of thermal noise, this system exhibits rich phase behaviours, including a
fluid state with diffusive dynamics, a cluster state, a hexatic state, a glassy state, a plastic crystal state and phase demixing.We reveal that these behaviours are induced by the off-axis
and many-body nature of nonlinear hydrodynamic interactions and the finite time required for propagating the interactions by momentum diffusion.
}
\end{abstract}

\maketitle

Self-organization is the autonomous organization of components into patterns or 
structures and the underlying mechanism can be grouped into two classes: thermal 
and athermal origin \cite{whitesides2002self,grzybowski2009self}. 
Compared to the understanding of self-assembly of a thermal system, which can be explained 
in terms of the free energy, that of an athermal system is 
far behind due to the lack of a firm theoretical background.  
Typical examples for the latter is active matter in hydrodynamic environments, 
which shows many interesting unconventional 
pattern formation with exotic dynamical features \cite{Marchetti2013,whitesides2002self,grzybowski2000dynamic,ishikawa2006interaction,lauga2009hydrodynamics,baskaran2009statistical,drescher2009dancing,ramaswamy2010mechanics,wensink2012meso,Stark2014}. 
Directional swimmers undergoing translational motion, like fishes, tend to swim together 
while forming a cluster of particular shape \cite{lauga2009hydrodynamics,baskaran2009statistical,ramaswamy2010mechanics}. 
The hexatic ordering of directionally self-propelling particles was also recently studied \cite{bialke2012crystallization}. 

Rotation is another important type of motion. Self-organization of passive rotors 
was first observed in laboratory experiments by Grzybowski, Stone, and Whitesides \cite{grzybowski2000dynamic}. 
Such rotation can be induced by applying a torque to a particle by magnetic \cite{grzybowski2000dynamic,grzybowski2001dynamic,grzybowski2002directed} 
and optical fields \cite{friese1998optical}. 
Self-organization of active rotors has also been studied intensively.  
For example, it was found that there are interesting stable bound states of spinning Volvox algae \cite{drescher2009dancing}. 
Furthermore, it was shown \cite{riedel2005self} that micro-organisms, like spermatozoa, self-organize into dynamic vortices and 
they form an array with local hexagonal order. 
This study indicates that large-scale coordination of cells can be regulated hydrodynamically, 
and chemical signals are not required. 
It was also shown recently that self-assembly of rotors is a generic feature of aggregating swimmers \cite{schwarz2012phase}. 
There has also been a theoretical prediction for 
intriguing self-organization of rotating molecular motors in membranes \cite{lenz2003membranes}. 
This and related problems have also been investigated by numerical simulations \cite{gehrig2006nonlinear,llopis2008hydrodynamic,leoni2010dynamics,goetze2010flow,gotze2011dynamic,Gompper2014,Stark2014}. 

As described above, there exist two types of rotors \cite{fily2012cooperative}: 
Rotors driven by external torques are called passive rotors, 
while those that are internally driven are called active rotors and many such examples can be seen in biological systems. 
A realistic realization of a truly active system of self-rotors in biological systems  
may be one in which the particles are torque dipole 
with no resultant net torque on the system \cite{lenz2003membranes,leoni2010dynamics,fily2012cooperative}. 
To elucidate the physics of self-organization of flow created by rotors at a finite Reynolds number, 
a system of hard disks each of which is rotated by an externally applied torque is an ideal model system. 
Ordering of this system  may be regarded as a nonequilibrium counterpart of thermodaynamic ordering of hard disks. 
We note that , unlike a thermodynamic system, where a state is selected solely by free energy, dynamic factors such as 
hydrodynamic interactions also affects the selection of a state of an out-of-equilibrium system.     

For example, it has been now recognized that nonlinear hydrodynamic interactions play 
crucial roles  in self-organization of rotating disks \cite{grzybowski2000dynamic,grzybowski2001dynamic,grzybowski2002directed}.  
In other words, the phenomena are beyond Stokes approximations, and may even have a 
link to self-organization of vortices  \cite{eyink2006onsager}, which is observed at a high Reynolds number. 
Vortex crystals are very interesting such examples \cite{durkin2000experiments,aref2003vortex}. 
The nonequilibrium, nonlinear, nonlocal, non-instantaneous nature of hydrodynamic interactions makes analytical approaches to this problem very difficult and thus 
numerical simulations are expected to play a crucial role. 
This problem also has a link to self-organization of a point vortex system \cite{rasmussen2002dynamics,aref2007point}, but the finite size 
and solidity of disks lead to far more rich behaviour. 
Here we employ a fluid particle dynamics (FPD) method \cite{FPD}, which we developed for studying 
hydrodynamic interactions between colloidal particles (Methods). This method treats 
a solid colloid as an undeformable fluid particle inside which the viscosity is 
considerably higher than the surrounding liquid. 
This approximation makes us free from solid-fluid boundary conditions, which significantly simplifies the computation.  
This method can quite naturally 
deal with many-body hydrodynamic interactions even at a high Reynolds number. 

In this Communication, we study self-organization of rotating hard disks 
in a two-dimensional (2D) incompressible liquid by using the FPD method. 
Like the Ising model for magnetic ordering or the hard sphere system for crystallization, 
this system may serve as a fundamental model system for studying 
dynamical phase behaviour caused by hydrodynamic interactions between rotating particles. 
The situation is similar to the above-mentioned experimental and theoretical works \cite{grzybowski2000dynamic,grzybowski2001dynamic,grzybowski2002directed}.  
We make our simulation in two dimensions (2D) to compare with the behaviour of its thermal counterpart, 2D hard disks, whose thermodynamic behaviour is 
reasonably understood \cite{NelsonB}. 
Here we report surprisingly rich phase ordering behaviours, such as aggregation, re-entrant 
order-disorder transition, glass transition, plastic crystal formation, and phase demixing in this class of strongly nonequilibrium systems. 

\section*{Results}

\subsection*{Behaviours of a single and a pair of rotating disks}  
\begin{figure}[!h]
\centering
\includegraphics[width=8.5cm]{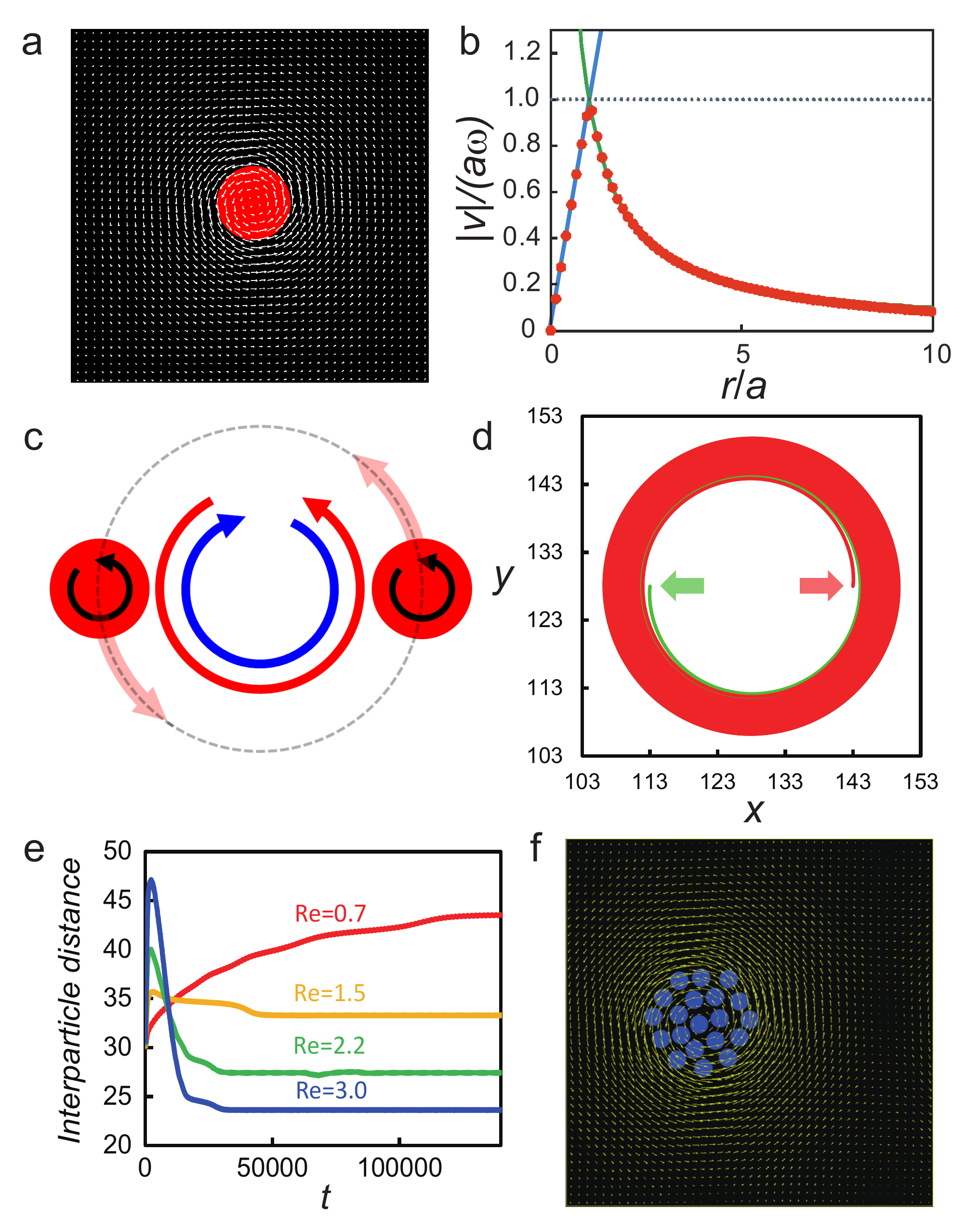}
 % fig1.eps: 0x0 pixel, 300dpi, 0.00x0.00 cm, bb=(atend)
\caption{{\bf Behaviours of an isolated rotating disk and co-rotating disks in a dilute suspension.} 
{\bf a,} 2D flow fields around an isolated rotating disk. 
{\bf b,} The velocity profile as a function of $r/a$. $v/(a \omega)$ increases linearly with $r$ 
inside the disk and then decays as $1/r$ in its outside, which is the characteristic of the so-called Rankin voltex. 
{\bf c,} Schematic picture of a pair of two disks rotating around the centre of mass. The two disks 
are rotating counter-clockwise with the same speed. 
{\bf d,} Trajectory of coupled rotating particles (the system size=256$\times$256). 
The initial separation of the particles (arrowed) was 30 and each particle rotates counter-clockwise with  $Re=0.76$. 
{\bf e,} Temporal change in the interparticle separation. We note that for all the cases including $Re=0.7$, 
the interparticle distance eventually reaches a final steady-state value of the separation (see Fig. \ref{fig:fig1}d), which monotonically decreases with $Re$. 
{\bf f,} A cluster of disks formed at $\Phi=0.04$ at $Re=5.94$. We confirm that irrespective of the value of $Re$, 
a cluster is always formed in the range of $Re$ studied. 
}
 \label{fig:fig1}
\end{figure}

Before discussing many-body interactions between disks rotating with an angular frequency $\Omega$,   
first we describe the behaviour of a dilute limit: Behaviours of a single rotating disk
and a pair of rotating disks (Supplementary Fig. 1 and Supplementary Note 1). 
We characterize the rotation speed of a disk either by the angular frequency $\Omega$ or the 
relevant Reynolds number $Re=\rho a^2 \Omega/\eta_\ell$ ($\rho$: density; $a$: particle radius: $\eta_\ell$: liquid viscosity). 
Figure \ref{fig:fig1}a and b show the 2D flow field around a rotating disk and the velocity distribution, respectively. 
The latter shows that the rotating velocity linearly increases with the distance from the centre of mass, $r$, inside a disk and 
decays as $1/r$ in its outside. This is the characteristic of the so-called Rankin vortex, i.e.,  
a forced vortex in the central core surrounded by a free vortex. 
We note that the Rankin vortex is known to mimic the 2D flow field of tornado and hurricane   \cite{montgomery2002experimental}.     
For two co-rotating `point' vortexes, the analytical solution is known and 
the two particles rotate around their center of mass towards the same direction as 
the rotating direction \cite{rasmussen2002dynamics,aref2007point}. In this case, the radius of rotation is the half of the initial interparticle distance. 
For particles with a finite size, on the other hand, the direction of rotation and the rotation center are the same 
as the case of point particles, 
but the radius of the rotation decreases with an increase in the rotation speed of the particles $\Omega$, or $Re$. 
This problem has a similarity to viscous interactions of co-rotating vortices \cite{le2002viscous,rasmussen2002dynamics}, but the 
crucial difference arises from the fact that the vortex core is an undeformable solid and not a liquid in our case. 
There are a few studies on spinning particles in three dimensions \cite{yeo2007dynamic,fily2012cooperative}; however, 
we note that there is an essential difference between 2D and 3D problems (Supplementary Figure 2 and Supplenetary Note 2). 
We confirm that if we fix the centres of mass of two rotating particles on a fixed line, they always repel 
with each other by the Magnus force \cite{landau1987fluid,grzybowski2000dynamic,grzybowski2001dynamic}. The attraction between particles is thus due to interparticle hydrodynamic 
interactions, as schematically explained in Fig. \ref{fig:fig1}c. 
The flow field generated by a rotating disk makes the other disk follow it, and vice versa. 
Thus, two particles try to follow each other while rotating around the centre of mass of the pair.  
In 3D, on the other hand, such hydrodynamic interactions are weaken due to the presence of the escape dimension, thus  
the Magnus force wins over the hydrodynamic attractive force, and particles repel with each other.  
In 2D, this hydrodynamic attraction overwhelms the repulsion due to the Magnus force,  
which leads to the rotating pair of particles (Fig. \ref{fig:fig1}d), whose interparticle distance monotonically decreases with an increase in $\Omega$, or $Re$ (Fig. \ref{fig:fig1} e), 
as long as the area fraction of disks, $\Phi$, is sufficiently small.  However, it should be noted that this situation is realized in a periodic boundary condition. 
In relation to the above dimensionality effect on hydrodynamic interactions between rotors, it is worth noting that the hexagonal ordering observed by Grzybowski et al.\cite{grzybowski2000dynamic,grzybowski2001dynamic,whitesides2002self,grzybowski2002directed} at finite Reynolds numbers is due to the effective hydrodynamic repulsion between the disks floating at an interface in three dimensions ({\it i.e.}, 2.5 dimensions).  

\subsection*{Structural ordering due to many-body hydrodynamic interactions}
\begin{figure}[!h]
\centering
\includegraphics[width=8.5cm]{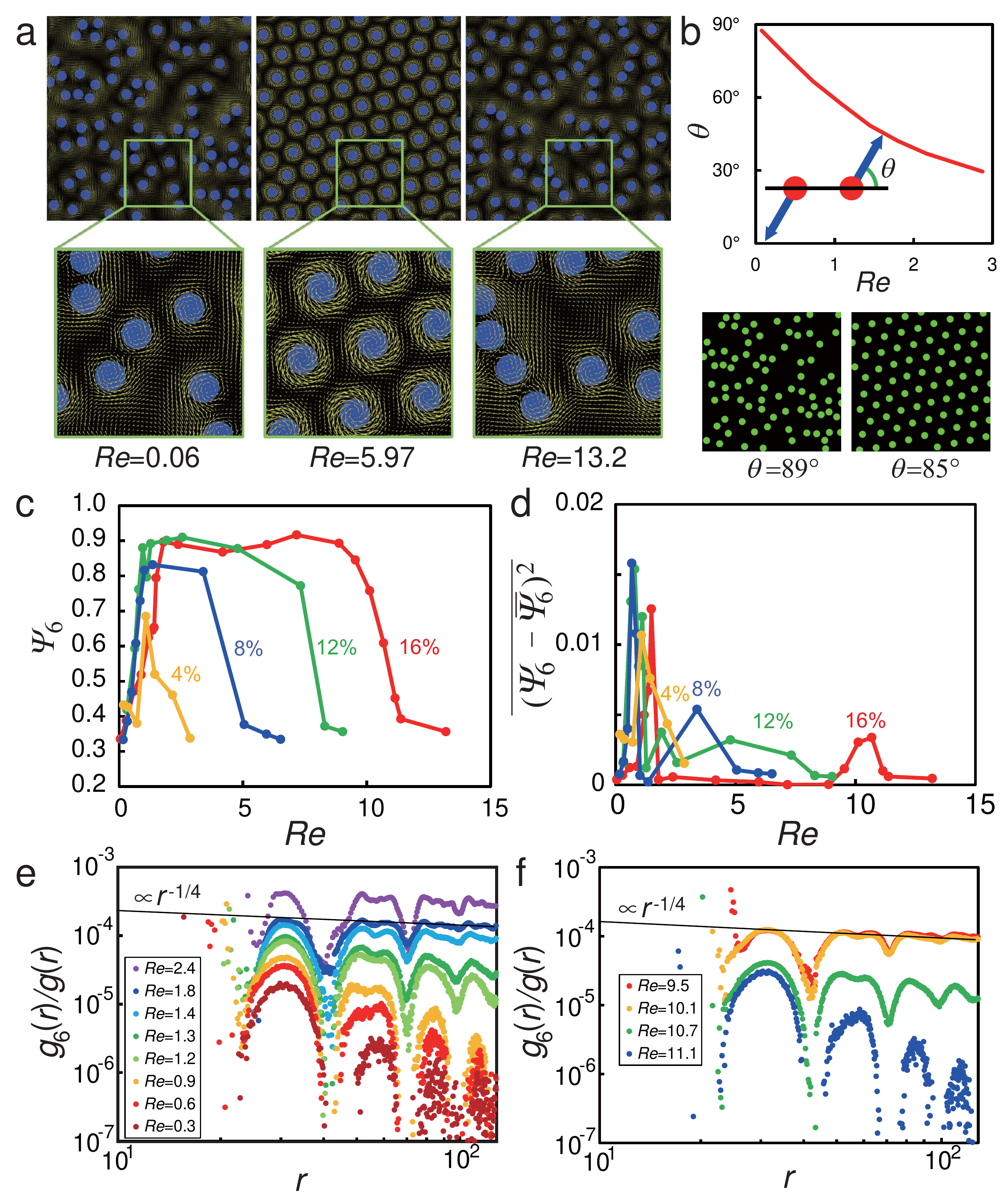}
 % fig1.eps: 0x0 pixel, 300dpi, 0.00x0.00 cm, bb=(atend)
\caption{{\bf Re-entrant state transitions observed at a rather high $\Phi$.}
{\bf a,} Snapshots of particle configurations together with the velocity fields. 
The system size is 256$\times$256, the number of particles is 80, and $\Phi \sim 0.157$. 
We can see sequential transitions from disordered, hexatic ordered, to disordered states with an increase in $Re$. 
{\bf b,} Upper panel: The direction of the hydrodynamic force $\theta$ measured from the interparticle axis. 
Lower panels: Patterns formed by Brownian simulations for the two off-axis forces: Left is a liquid state for $\theta=89^\circ$, whereas 
the right panel is a hexatic state for $\theta=85^\circ$. 
{\bf c,} $Re$-dependence of the hexatic order parameter $\Psi_6$, which clearly shows the re-entrant behaviour for various $\Phi$'s (expressed in \%). 
{\bf d,} The degree of fluctuations of $\Psi_6$ as a function of $Re$ for various $\Phi$'s (expressed in \%). We can see that as in a thermodynamic hexatic ordering transition, 
the order parameter exhibits large amplitude fluctuations near the transition points. 
{\bf e,} $Re$-dependence of the decay of the spatial correlation function of the hexatic order parameter $g_6(r)$ normalized by the radial distribution function $g(r)$ 
around the transition at low $Re$ for $\Phi=0.157$. 
In the hexatic state, $g_6(r)/g(r)$ decays with a power law with the exponent of -1/4, as it should be \cite{NelsonB}.  
In the disordered states it decays almost exponentially. 
{\bf f,} The same as {\bf e} but for around the transition at high $Re$. 
In the hexatic state, $g_6(r)/g(r)$ again decays with a power law with the exponent of -1/4.  
In the disordered states it decays almost exponentially. 
}
 \label{fig:fig2}
\end{figure}

Now we consider the dynamical behaviour of a system of many disks rotating in the counter-clockwise 
direction with $\Omega$. 
The hydrodynamic attraction between rotating disks leads to the formation of rotating clusters for low $\Phi$ (Fig. \ref{fig:fig1}f). 
The higher rotation speed of individual disks  leads to the formation of a more compact rotating cluster.  
This tendency is basically the same as that for a pair of rotors (Fig. \ref{fig:fig1}e). 
For this regime, only one cluster is formed in the simulation box and the whole cluster rotates in the counter-clockwise 
direction, as shown in Fig. \ref{fig:fig1}f. A cluster always tends to have a circular shape, but it does not have any particular internal structural order, 
partly because imperfect matching between the size and the number of particles leads to structural fluctuations:   
the internal structure is basically controlled by the number of disks in it and $Re$, which determine the cluster size, but fluctuating with time.   
With a further increase in $Re$, however, this cluster state becomes unstable, since  
the repulsive Magnus force of nonlinear origin eventually wins over the hydrodynamic attraction for high $Re$ (see below).  
Above  a critical $\Phi$  ($\Phi_c \sim 0.05$), on the other hand, a system exhibits a re-entrant transition between states  
as a function of $\Omega$ (Fig. \ref{fig:fig2}a): for low $\Omega$ a system is in a disordered liquid state with large fluctuations, 
but with an increase in $\Omega$ it enters into a rather stable hexatic phase where rotating particles 
are localized on a hexagonal lattice. 
The border between the cluster to the hexatic state is rather sharp as a function of $\Phi$. 

\begin{figure}[htbp]
  \begin{center}
   \includegraphics[width =8.5cm]{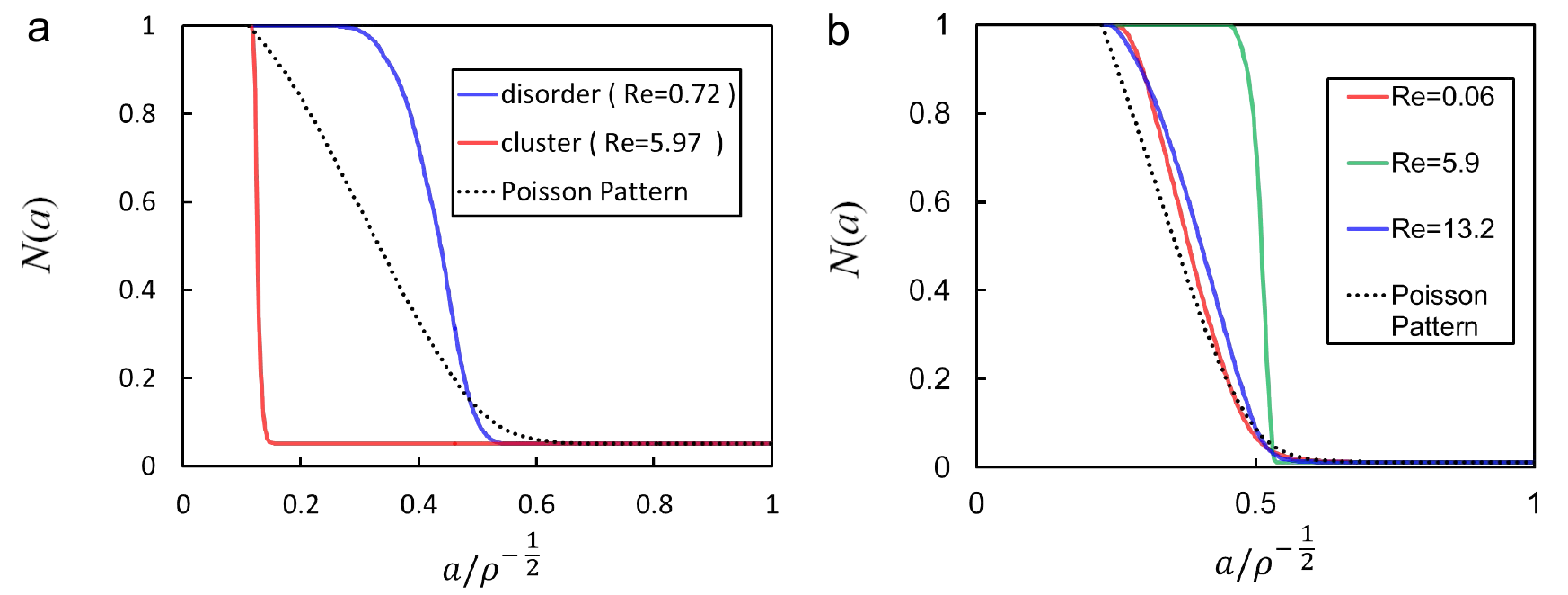}
   \end{center}
  \caption{{\bf The analysis of point patterns.} 
  {\bf a,} The decrease of $N$ as a function of the particle radius $a/\rho^{1/2}$ for $\Phi=0.04$, where 
  $\rho$ is the particle number density and $\rho^{-1/2}$ is the average interparticle distance. 
{\bf b,} The decrease of $N$ as a function of the particle radius $a/\rho^{1/2}$ for $\Phi=0.15$.   }
  \label{fig:fig3}
\end{figure}

The transition can be characterized by the nature of interparticle interactions. 
Here we analyse a point pattern to extract the nature of interparticle interactions. 
The point pattern analysis is very useful to determine the overall interparticle interactions \cite{tanaka1989digital}. 
Here we use what we call $N$ function, which is the number of connected regions as a function 
of the radius of circle whose centre is located at the centre of each disk. The number of connected regions 
decreases monotonically, with an increase in the circle radius, from the total number of disks $N_0$ 
to one. Here we use $N$ for the one normalized by $N_0$. 
By comparing $N$ for a reference system made of randomly distributed disks, i.e., Poisson pattern, 
we can judge whether the interaction is repulsive or attractive. For a system of particles with repulsive 
interactions, particles form a rather regular pattern and $N$ decays slower than that for the corresponding Poisson pattern. 
For a system of particles with attractive 
interactions, on the other hand, particles form a cluster pattern and $N$ decays faster than that for the corresponding Poisson pattern. 
In Fig. \ref{fig:fig3}a, we show the $N$ function for $\Phi=0.04$. We can clearly see that for the cluster-forming case at $Re=0.72$ the interaction is attractive 
whereas for the disordered state at $Re=5.97$ the interaction is repulsive. 
In Fig. \ref{fig:fig3}b, we show the $N$ function for $\Phi=0.15$.
We can see that for all the value of $Re$ the interaction is basically repulsive, but 
its strength is maximum at $Re=5.9$, for which the hexatic order is formed.

The transition can be seen even more clearly by the degree of localization of the flow field: 
for a hexatic state each particle has its own localized rotational flow field, whereas for a cluster state a single vortex is always formed and thus the flow field 
is strongly delocalized. 
However, the precise nature of the cluster-hexatic phase transition, such as whether the transition is continuous or discontinuous and  
whether a disorder state always exists between the two states or there exists a critical point between the cluster and the hexatic state, 
is not clear at this moment. 
To access this problem, we need to survey the border region of a bigger system size with a high resolution of $\Phi$ and $Re$. 
Although this is a very interesting problem, we leave this for future investigation. 

\begin{figure}
\begin{center}
\includegraphics[width =6cm]{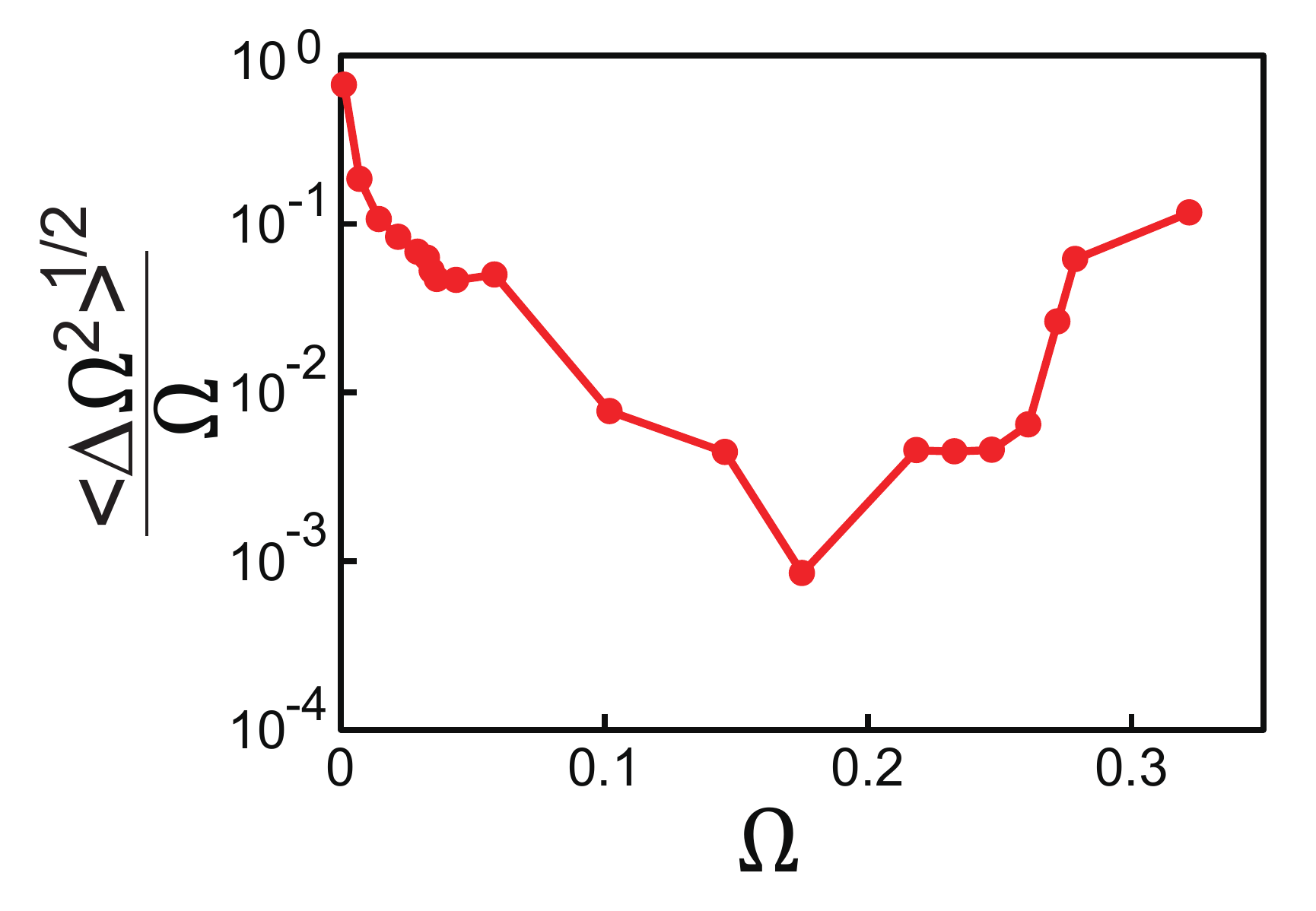}
\end{center}
\noindent
\caption{{\bf The normalized variance of the angular frequency $\Omega$ 
of rotating disks as a function of the averaged $\Omega$ for $\Phi=0.15$.} 
We can see the negative correlation between the degree of the hexatic order and the variance. }
\label{fig:fig4}
\end{figure}

We also find that the hexatic state is eventually destabilized and melts by a further increase in $\Omega$ 
and a system becomes disordered again.  
We note that for all these states interparticle interactions are basically repulsive, unlike the case of low $\Phi$. 
Although the torque exerted to each disk is exactly the same, the rotation speed of a disk can in principle depend 
on the particle configuration around it. Here we show in  Fig. \ref{fig:fig4} the normalized variance of the angular frequency $\Omega$ 
of rotating disks as a function of the averaged $\Omega$. For ordered states, the variance, i.e., the 
fluctuations of rotation speed, becomes very small, indicating all the disks rotating with almost the same frequency. 
For disordered states, on the other hand, the variance is large, reflecting the large fluctuations of particle environment. 
This result indicates a strong negative correlation between the degree of fluctuations of rotational speed of particles and the degree of order

In Fig. \ref{fig:fig2}c, we show the $\Omega$-dependence of the hexatic order parameter $\Psi_6$, which clearly indicates 
the re-entrant nature of the state transitions. The transitions can also be characterized by 
the magnitude of fluctuations of $\Psi_6$, or the susceptibility (see Fig. \ref{fig:fig2}d), which is also observed in a thermodynamic hexatic 
ordering in 2D disks. 
In the hexatic ordered state, we confirm the power law decay of the spatial correlation 
of the hexatic order which is specific to the hexatic phase (see Fig. \ref{fig:fig2}e and f). 
We do not see any indication of the positional order since the radial distribution function 
decays almost exponentially (see below). 

\begin{figure}[h!]
  \begin{center}
   \includegraphics[width =8.5cm]{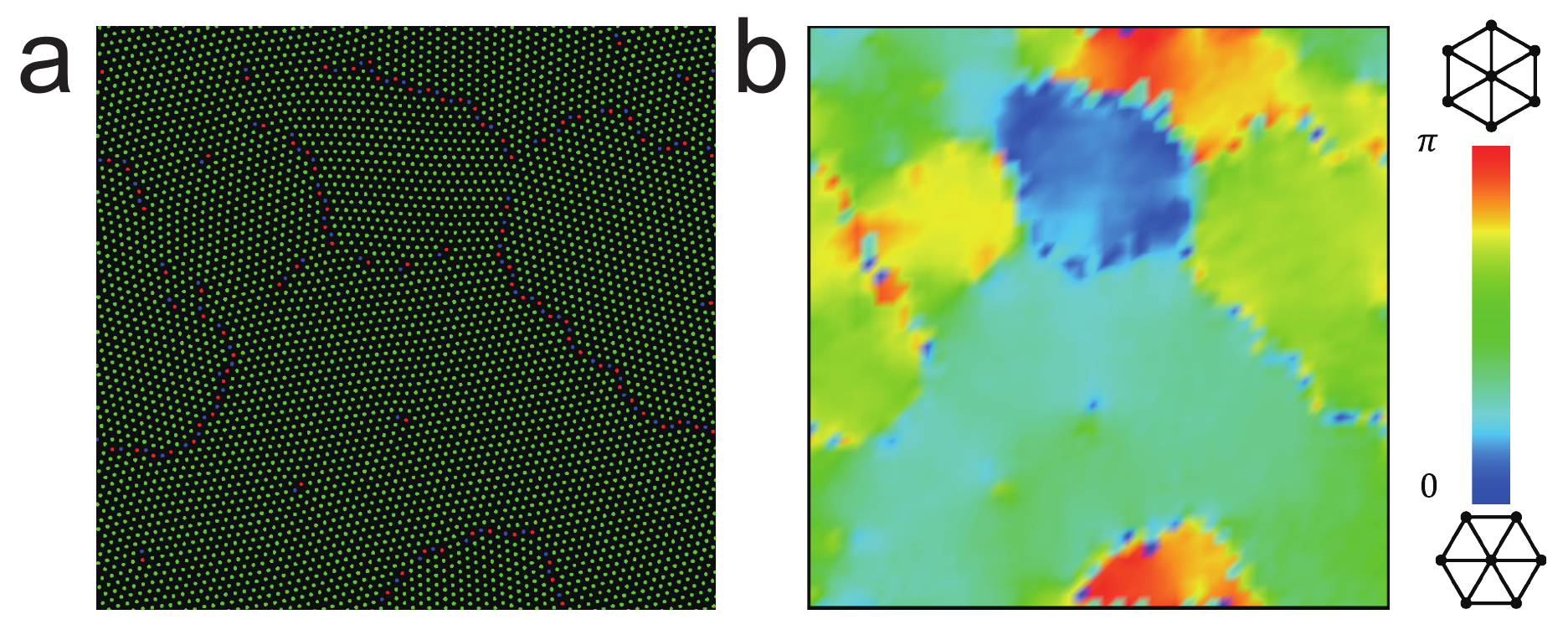}
  \end{center}
  \caption{{\bf Hexatic ordering in a large system.} 
  Here we show results of large-size simulation (lattice size=2048$^2$ and the number of particles=5120). 
  The area fraction $\Phi=0.15$ and $Re=5.9$. 
  {\bf a,} The structure of a hexatic state shown with three colours (yellow for particles with six neighbours; red for particles with more than 7 neighbours; 
  blue for particles with less than 5 neighbours).  
  {\bf b,} The same as {\bf a} but with orientations of hexatic order. See the colour bar on the meaning of colour. 
  }
  \label{fig:fig5}
\end{figure}

\begin{figure}[h!]
  \begin{center}
   \includegraphics[width =8.5cm]{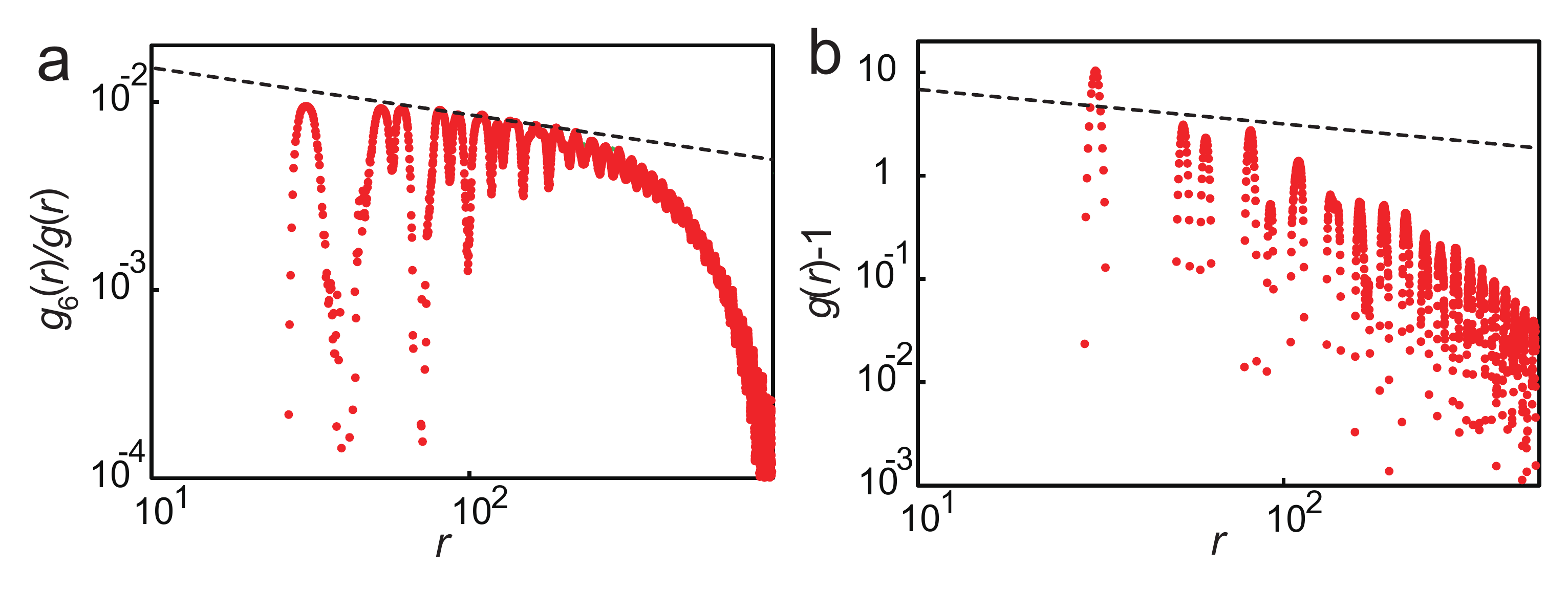}
  \end{center}
  \caption{{\bf Spatial correlation of hexatic and positional order.} 
  {\bf a,} The spatial decay of the correlation function of the hexatic order $g_6(r)$ in Fig. \ref{fig:fig5}a normalized by 
  the radial distribution function $g(r)$. We can see it decays algebraically in a short distance, but decays faster for long distance 
  because of the finiteness of the domain size. The dashed line has a slope of $-1/4$. {\bf d,} The spatial decay of the radial distribution function $g(r)$ 
  for the pattern shown in Fig. \ref{fig:fig5}a. 
  It decays more quickly than the decay of $g_6(r)$ even for rather short distance ($r<400$). The dashed line has a slope of $-1/3$.
  }
  \label{fig:fig6}
\end{figure}

Here we show a hexatic phase observed in a large system (lattice size=2048$^2$ and the number of particles=5120) 
at $\Phi=0.15$ and $Re=5.9$.  
We can see grain boundaries between hexatic order with different orientations in Fig.  \ref{fig:fig5}a and b. 
Figure \ref{fig:fig6}a and b show the decay of the correlation function of the hexatic order normalized by 
the radial distribution function $g(r)$, $g_6(r)/g(r)$, and that of $g(r)-1$, respectively. 
We can see $g_6(r)/g(r)$ decays algebraically in a short distance $r<400$, but decays faster for long distance. 
This is because the size of mono-domain regions is finite (see Fig.  \ref{fig:fig5}a and b). 
On the other hand, we can see that $g(r)$ decays faster than an algebraic decay even for $r<400$, where $g_6(r)/g(r)$ decays algebraically. 
This clearly indicates the absence of quasi-long-range translational order in the ordered phase.  
Thus we conclude that the ordered phase is the hexatic phase.  
The appearance of the transitions between dynamical states as a function of $\Phi$ and $\Omega$ 
in athermal systems is quite striking, which is reminiscent of phase transitions in a thermal system. 
We stress that the interparticle interaction in our system is of purely hydrodynamic origin. 

Here we consider a quite interesting feature of hydrodynamic interparticle interactions, which are 
absent in a thermal system. Binary interaction potentials that lead to phase ordering in thermal systems 
always act along the interparticle axis, i.e., along the line connecting the centres of mass of two interacting particles. 
However, this is not the case for our athermal system. The nonlinear Magnus force acts along the 
line connecting two particles, whereas the Stokes force acts according to the flow direction. 
Furthermore, linear hydrodynamic interactions are tensorial and act not along the 
interparticle direction. Thus, the total hydrodynamic force does not act along the 
interparticle axis. 
The direction of the hydrodynamic force as a function of $Re$ is shown in the upper panel of Fig. \ref{fig:fig2}b with a 
schematic explanation. 
With an increase in $Re$, both the Magnus and the hydrodynamic force increase.  
However, the nonlinear Magnus force increases 
more rapidly than the linear hydrodynamic force. Thus, the direction of the total force, which is repulsive, approaches the interparticle 
axis. A strong enough repulsive force acting on particles not so far from the interparticle direction leads to the formation 
of hexatic order. To verify this scenario, we have performed Brownian dynamics simulation (without hydrodynamic interactions) by changing the angle $\theta$ 
between the direction of an artificial repulsive force and the interparticle direction. We find that when $\theta$ is continuously decreased, 
a system indeed forms hexatic order below a critical value of $\theta$ (see the lower panels of Fig. \ref{fig:fig2}b). 

\begin{figure*}[!t]
 \centering
 \includegraphics[width=11cm]{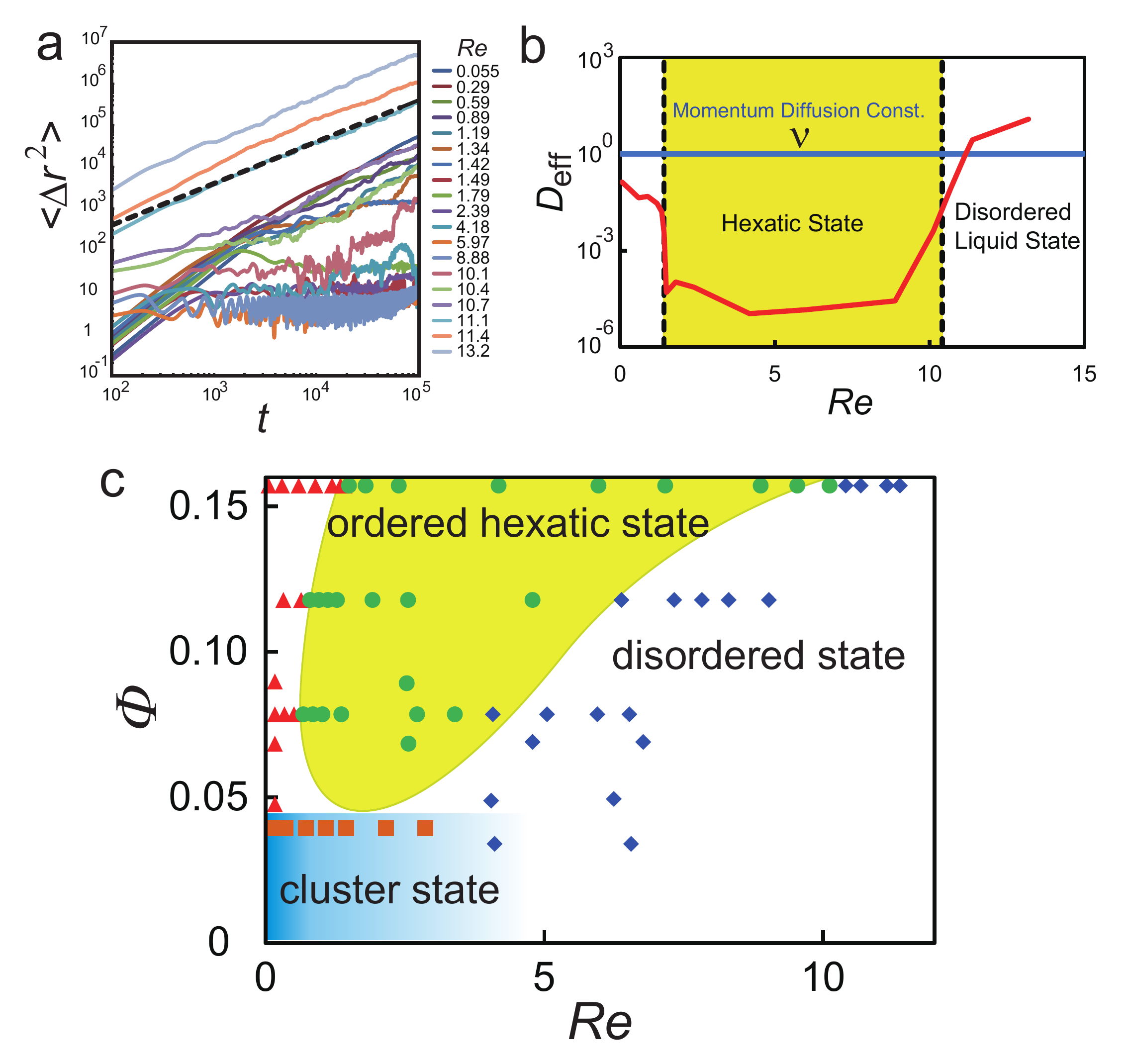}
 % fig1.eps: 0x0 pixel, 300dpi, 0.00x0.00 cm, bb=(atend)
\caption{{\bf Dynamical behaviour and state diagram of rotating disks. }
{\bf a,} The mean-square displacement $\langle \Delta r^2 \rangle^{1/2}$ vs. time $t$. 
Despite the absence of thermal noise, particles undergo Brownian-like diffusive motion $\langle \Delta r^2 \rangle^{1/2}=D_{\rm eff} t$ 
with an effective diffusion constant $D_{\rm eff}$ in the long-time limit. 
The black dashed line is the momentum diffusion constant, or the kinematic viscosity $\nu$($=\eta_\ell/\rho$). 
{\bf b,} $Re$-dependence of the effective diffusion constant $D_{\rm eff}$. In the hexatic ordered state, 
$D_{\rm eff}$ is very low and there is no diffusive behaviour, as it should be (see {\bf a}). 
At high $Re$, $D_{\rm eff}$ exceeds $\nu$ (indicated by the blue horizontal line), which means that fluctuations of particles cannot be suppressed by hydrodynamic interactions at high $Re$ 
(see text). 
{\bf c,} State diagram on the $Re$-$\Phi$ plane. At low $\Phi$ and low $Re$, the system forms a cluster due to hydrodynamic attractive interactions. 
At high $\Phi$, on the other hand, the system exhibits re-entrant transition between a disordered chaotic liquid state 
and ordered hexatic state. The latter state is basically stabilized by the repulsive interaction due to the Magnus effect. 
}
 \label{fig:fig7}
\end{figure*}

Next we discuss the nature of the translational motion of rotating disks in the disordered states. 
To see this, we calculate the mean-square displacement of disks $\langle \Delta r^2(t) \rangle$ (see Fig. \ref{fig:fig7}a). 
Interestingly, in the two types of disordered states particle motion is apparently diffusional: 
In the long-time limit we observe the relation 
$\langle \Delta r^2(t) \rangle \sim D_{\rm eff}t$, where $D_{\rm eff}$ is the effective diffusion constant and $t$ is the time duration. 
It should be noted that in our system there is no thermal noise and an isolated rotating disk 
exhibits no motion. 
For a pair of rotating particles, we also observe their trajectory is very stable and there is no fluctuation (see Fig. \ref{fig:fig1}d). 
Thus, the apparently diffusional motion should be the consequence of self-generated force noise due to many-body hydrodynamic 
interactions of both linear and nonlinear origin. We can also see no diffusion, or non-ergodic behaviour, for the hexatic state. 
The random nature of fluctuations may be related to the long-range nature of hydrodynamic interactions, which 
makes a number of the surrounding particles affecting a particle large enough to provide strong \emph{stochastic} spatio-temporal 
fluctuations. 

For high $Re$, the particle diffusion starts to become comparable or faster than momentum diffusion: $D_{\rm eff} \geq \nu$, where 
$\nu$ is the momentum diffusion constant or the kinematic viscosity $\nu=\eta/\rho$ (see Fig. \ref{fig:fig7}a and b). 
This implies that hydrodynamic interactions induced by the rotation of a particle cannot fully propagate to its neighbouring particles for high $Re$. 
This weakens the repulsive interactions of particles and eventually leads to the melting of the hexatic state. 
Thus the re-entrant hexatic ordering as a function of $Re$ may be explained as follows: 
The increase of the Magnus force of nonlinear origin with an increase in $Re$ makes the direction of the interparticle force 
more aligned along the interparticle direction and also increases the strength of the repulsive force. 
This leads to stabilization of the hexatic order, as explained above. The ordered state is stable until the repulsive interaction 
is weakened by the intrinsically kinetic nature of hydrodynamic interactions: Unlike ordinary interparticle 
interactions which propagate with the speed of light, hydrodynamic interactions propagate much slower in a diffusive 
manner with the momentum diffusion constant $\nu$. This kinetic weakening of the repulsive force eventually destabilizes 
the hexatic state and leads to the melting into the disordered chaotic state. 

Here we summarize what we observed in our system and show the state diagram as a function of $\Phi$ and $Re$ (see Fig. \ref{fig:fig7}c). 
We can see the three states, i.e., disordered fluid, cluster, and ordered hexatic state. 
It is quite striking that particles interacting by hydrodynamic interactions alone exhibit such rich phase 
behaviours.   

\begin{figure*}[t!]
 \centering
 \includegraphics[width=11cm]{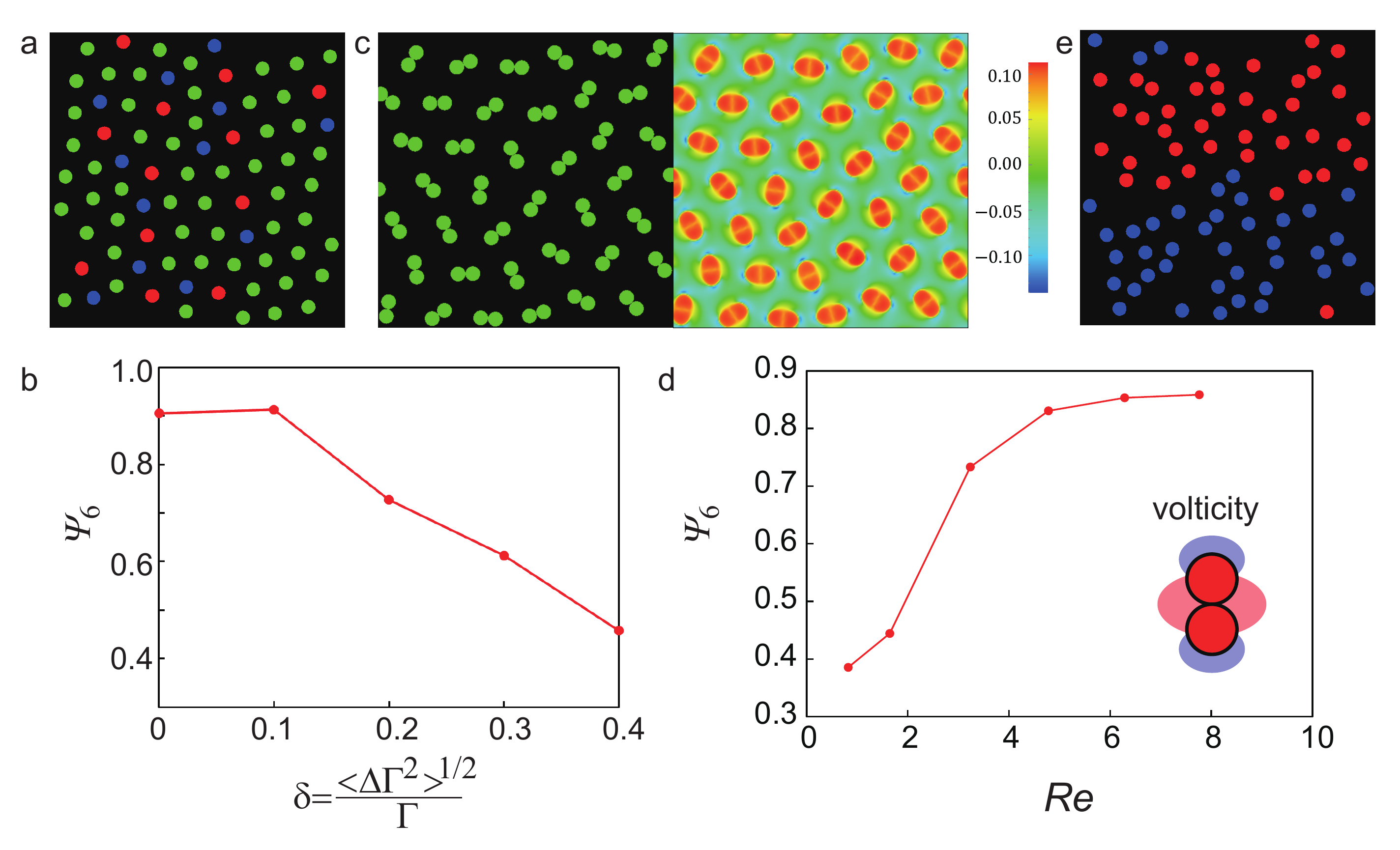}
 % fig1.eps: 0x0 pixel, 300dpi, 0.00x0.00 cm, bb=(atend)
\caption{{\bf Other interesting states formed in a system made of rotating particles. }
{\bf a,} A glassy non-ergodic states formed in a system of rotating disks where the torque $\Gamma$ acting on particles 
has a Gaussian distribution whose normalized variance is $\delta=\langle \Delta \Gamma^2 \rangle^{1/2}/\Gamma$. 
Here $\Phi=0.157$, the average $Re$ ($\bar{Re})=4.22$, and $\delta=0.2$. 
The colour of particles is green when the number of nearest neighbour particles $NN$ is 6. 
For $NN>6$, the colour is red and for $NN<6$ blue.  
{\bf b,} $\delta$-dependence of the hexatic order parameter $\Psi_6$. 
Here $\Phi=0.157$ and $Re=7.76$. 
With an increase in $\delta$, 
the hexatic order monotonically decreases and the system eventually enters into a nonergodic glassy state. 
{\bf c,} A plastic phase formed in a system made of rotating dumbbells (counter-clockwise). The torque applied is the same for all particles. 
The rotation direction is counter-clockwise. The system exhibits hexatic order at a certain range of $Re$, but without any 
orientational order of the axes of dumbbells.  
{\bf d,} $Re$-dependence of $\Psi_6$ for rotating dumbbells. 
We note that the transition is rather broad. The schematic image represents the distribution 
of volticity around a single rotating dumbbell (red: counter-clockwies; blue:clockwise). 
{\bf e,} Phase separation of particles rotating clockwise ($\Omega$) and counter-clockwise ($-\Omega$) ($\Omega=0.155$ (or, $Re = 6.35$)). 
$\Phi=0.157$ and the number fraction of particles rotating clockwise is 0.5. 
}
\label{fig:fig8}
\end{figure*}

\subsection*{Other interesting states formed by rotating disks}
\begin{figure}
  \begin{center}
   \includegraphics[width =8.5cm]{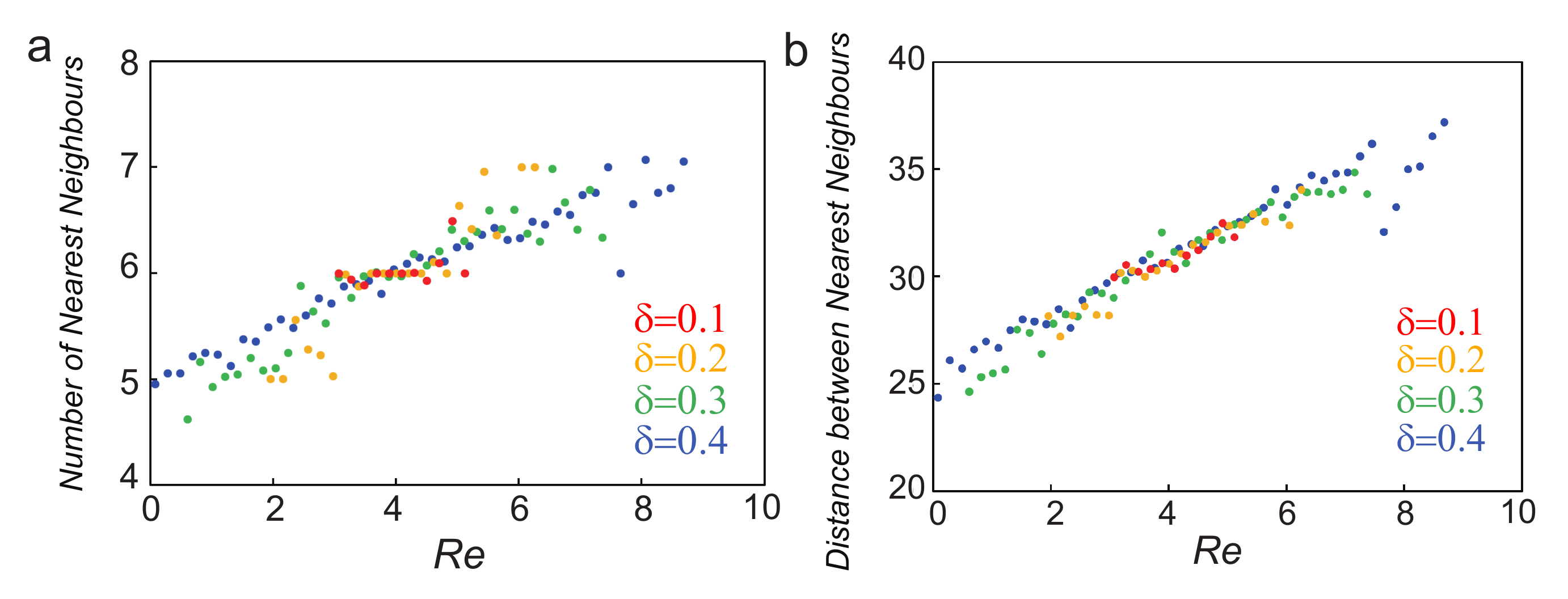}
  \end{center}
\caption{{\bf Correlation between local structure and rotation speed (or, Re), of each particle in a glassy state.} 
  {\bf a,} Dependence of the number of nearest neighbours on $Re$  for individual particles. The system size is 256$^2$, $\phi=0.15$, and the 
  average value of $Re$, $\bar{Re}$, is 4.2. We can see that the larger deviation from $\bar{Re}$ leads to the larger deviation from 
  the average number of nearest neighbours (=6).  
  {\bf b,} Dependence of the distance to nearest neighbour particles on $Re$ for individual particles. The conditions are the same as the above. 
  We can see that the the interparticle distance is larger for a particle rotating with a faster speed (or, larger Re) because of stronger hydrodynamic repulsive force, i.e., Magnus force. 
}
\label{fig:fig9}
\end{figure}

Finally, we show other interesting states formed by rotating particles. 
The introduction of size polydispersity to hard spheres is known to lead to the formation of a glass state for 
a thermal system \cite{KAT}. Motivated by this, we introduce the polydispersity in the rotating speed of particles, whose variance is $\delta$,  
and indeed find a non-ergodic glassy state of the rotating particles (Fig. \ref{fig:fig8}a) for high enough $\delta$ (Fig. \ref{fig:fig8}b): liquid-glass transition in an athermal system. 
Reflecting the nearly continuous nature of the liquid-to-hexatic transition \cite{NelsonB,krauth}, 
there is no sharp transition from a hexatic to a disordered glassy state.  
Although disks are rotating around their centres of mass, their  positions are frozen in a disordered configuration 
and thus the system can be regarded as a nonergodic glassy state.  
Here we show how the distribution of the rotation speed of particles leads to the loss of hexatic order and results 
in the formation of a glassy state. As shown in  Fig. \ref{fig:fig9}a, the larger deviation from the averaged $Re$ leads to the larger deviation from 
the average number of nearest neighbours (=6). With an increase in the variance of the distribution $\delta$, more defects are produced and the hexatic order 
is eventually lost above $\delta \geq 0.2$.  We can also see that the the interparticle distance is larger for a particle rotating with a faster speed because of stronger hydrodynamic repulsive force, i.e., Magnus force (Fig. \ref{fig:fig9}). Both of these disorder effects are responsible for the formation of a glassy state: 
The number of the nearest neighbours and the average distance to the neighbours 
are both strongly correlated to the rotational speed of particles, which explains a wide enough distribution of the rotation speed 
results in the formation of a non-ergodic amorphous state instead of a hexatic ordered state. 
This glassy state is non-ergodic if we consider the particle configuration, yet maintains strong flow fields, which makes this state very unique. 
Reflecting the kinetic origin of interparticle interactions, the introduction of disorder in the dynamic 
quantity, $\Omega$, is essential for avoiding the ordering, which is an interesting point unique to 
purely kinetic athermal systems.

We also find a plastic crystal-like state in a rotating particle pair (dumbbell) system (Fig. \ref{fig:fig8}c) above a critical $Re$ (Fig. \ref{fig:fig8}d). 
This state is characterized by hexatic ordering of the centre of mass of disk pairs (dumbbells) 
without any orientational order in the axis directions of dumbbells. 
So we find for this type of athermal systems made of rotating particles 
almost all states seen in its thermodynamic counterparts, including 
a liquid, a hexatic phase, a glass, and a plastic crystal. 
The only missing state is a liquid-crystalline state, but the lack of this state is natural 
consequence of the fact that the system is composed of rotating elements. 
We also note that the modification of the rotational direction leads to complex behaviours; for example, 
we can introduce phase demixing of particles rotating oppositely (Fig. \ref{fig:fig8}e), which may be used to separate 
different types of passive and active rotors. In relation to this, it is worth noting that phase separation between particles rotating clockwise and anti-clockwise was recently observed even without hydrodynamic interactions \cite{nguyen2013emergent}. 
It is interesting that such phase separation is observed in both systems with and without hydrodynamic interactions. 
It is also worth noting that the presence of regular arrays of vortices, e.g., the triangle state, 
are predicted for an active polar film \cite{voituriez2006generic}.  So far we have not seen such a regular state, but the similarity in the physics between the two systems 
implies that it might exist in a certain parameter range. This is an interesting problem for future study.

\section*{Discussion}

It is remarkable that our athermal system where particles interact only via hydrodynamic interactions 
exhibits such rich phase (or more strictly, state) behaviour and reproduces almost all the physical states observed in its thermal counterpart.  
We hope that these phase behaviours will be observed experimentally. 
For this purpose, a quasi-2D version of the experimental setup used in \cite{grzybowski2000dynamic,grzybowski2001dynamic,whitesides2002self,grzybowski2002directed} 
may be suitable, since the 2D nature of hydrodynamic interactions is important.  
Here we have focused on the bulk behaviour of rotating disks 
to study the fundamental nature of dynamical phase behaviour from a viewpoint of nonequilibrium statistical physics. 
The effects of confinement on such a system are also quite interesting not only from a fundamental viewpoint, 
but also from a viewpoint of applications to microfluidics \cite{terray2002microfluidic,bleil2006field}. Such effects on rotors in a fluid 
have recently been numerically studied by G\"otze and Gompper \cite{goetze2010flow, gotze2011dynamic}. 
It is quite interesting to study how such spatial confinements affect all the dynamical phases we reported and the transition between them.  

Here it is worth stressing that even non-ergodic states of particles such as hexatic, glassy, and plastic crystal states 
have strong hydrodynamic flow fields, which makes these states quite distinct from their thermal counterparts.  
The interesting and unique feature of hydrodynamic self-organization is that structural ordering is the consequence of {\it self-organization of flow} dissipating 
energy and thus even non-ergodic states are maintained by dynamical flow. 
In nature, there are many dynamical systems in which crucial interactions between elements 
are of purely hydrodynamic origin. 
The coexistence of linear and nonlinear hydrodynamic interactions, the resulting 
unconventional off-axis force, the finite propagation speed of the interactions, and the significance of hydrodynamic degrees of freedom 
even for non-ergodic (apparently static) states lead to rich and non-trivial self-organization.     
We hope that our study sheds new light on {\it hydrodynamic self-organization} and stimulate further study on this intriguing 
problem.  

%\clearpage
\vspace{3cm}
\noindent
{\bf METHODS}

\noindent
{\bf Simulation method.}
Treating hydrodynamic interactions between colloids is difficult even for a thermal system. 
There are several methods such as Stokesian dynamics, lattice-Boltzmann, smooth-particle, 
and fluid-particle-dynamics (FPD) methods. Here we employ our FPD method \cite{FPD,furukawa2010key}, which has 
an advantage in its theoretical transparency and its applicability to a high Reynolds number ($Re$) regime. 
We can access a high $Re$ regime rather easily particularly because a torque applied externally makes 
the flow field inside a fluid disk almost exactly that for a solid disk even at high $Re$.  

Here we briefly explain the FPD method \cite{FPD} and the physical concept behind it. 
A particle whose centre of mass is located at ${\mbox{\boldmath$r$}}_i$ 
is represented by a smooth viscosity change as 
$\eta({\mbox{\boldmath$r$}})=\eta_{\rm \ell}+\sum_{i}^{N}(\eta_{\rm c}
-\eta_{\rm \ell})\phi_i({\mbox{\boldmath$r$}})$,
where $\eta_{\rm \ell}$ is the liquid viscosity and 
$\eta_{\rm c}$ is the viscosity inside a colloid particle. 
The summation is taken over all $N$ particles. 
$\phi_i$ represents particle $i$ as 
$\phi_i({\mbox{\boldmath$r$}})=\{\tanh[(a-|{\mbox{\boldmath$r$}}
-{\mbox{\boldmath$r$}}_i|)/\xi]+1\}/2$, 
where $a$ is the particle radius and $\xi$ is the interface thickness.
Then the equation of motion to be solved is 
\begin{eqnarray}
\rho\bigl(\dfrac{\partial}{\partial t}
+\mbox{\boldmath$v$}\cdot\nabla\bigr)\mbox{\boldmath$v$}
&=& {\mbox{\boldmath$f$}_{U}}+{\mbox{\boldmath$f$}_{T}}-\nabla\cdot {\stackrel{\leftrightarrow}{\mbox{\boldmath$\Pi$}}} 
\label{Navier-Stokes}
\end{eqnarray}
with ${{\stackrel{\leftrightarrow}{\mbox{\boldmath$\Pi$}}}} = p{\stackrel{\leftrightarrow}{\mbox{\boldmath$I$}}}-\eta(\nabla{\mbox{\boldmath$v$}}^\dagger+\nabla{\mbox{\boldmath$v$}})$, 
where $\stackrel{\leftrightarrow}{\mbox{\boldmath$I$}}$ is the unit tensor. 
Here $\rho$ is the mass density, 
and we assume that the density of the liquid is the same as that of particles. 
$\mbox{\boldmath$v$}({\mbox{\boldmath$r$}})$ is the velocity field, 
and the pressure $p$ is determined to satisfy 
the incompressibility condition $\nabla\cdot{\mbox{\boldmath$v$}}=0$. 
Here ${\mbox{\boldmath$f$}_{U}}({\mbox{\boldmath$r$}})$ 
is the force density due to the interparticle interaction determined as 
${\mbox{\boldmath$f$}_U}({\mbox{\boldmath$r$}})= -\sum_{i}^{N}
(\phi_i({\mbox{\boldmath$r$}})/A)\sum_{j\ne i}^{N}\partial 
U(|{\mbox{\boldmath$r$}}_{ij}|)/\partial {\mbox{\boldmath$r$}}_{ij}$, 
where $U(r)$ is the interparticle potential, 
${\mbox{\boldmath$r$}}_{ij}={\mbox{\boldmath$r$}}_{i}
-{\mbox{\boldmath$r$}}_{j}$, and $A=A_{i}=
\int d{\mbox{\boldmath$r$}}\phi_i({\mbox{\boldmath$r$}})$ 
is the area of each particle.  
${\mbox{\boldmath$f$}_{T}}({\mbox{\boldmath$r$}})$ 
is the force density due to the torque:  
${\mbox{\boldmath$f$}_{T}}({\mbox{\boldmath$r$}})=\alpha |\mbox{\boldmath$r$}| \phi(\mbox{\boldmath$r$}) \mbox{\boldmath$e$}_\theta$, 
where $\mbox{\boldmath$e$}_\theta$ is the unit angular vector in the counter-clockwise direction. 
$\alpha$ is the strength of the torque and $\alpha>0$ leads to the counter-clockwise rotation of a particle.

In our FPD method the particle rigidity is approximately expressed by introducing the smooth viscosity profile, $\eta
(\mbox{\boldmath$r$})$. The approximation is better for a larger viscosity ratio $\eta_{\rm c}/\eta_{\rm \ell}$ 
and a smaller $\xi/a$. By multiplying both sides of Eq. (\ref{Navier-Stokes}) 
by $\phi_i({\mbox{\boldmath$r$}})$ and then performing its spatial integration, we can straightforwardly obtain 
an approximate equation of motion of particle $i$: $M_i d \mbox{\boldmath$V$}_i/d t={\mbox{\boldmath$F$}}_i
+{\mbox{\boldmath$K$}}_i$, where $M_i=\rho A=M$ and ${{\mbox{\boldmath$V$}}_i}=\int d {\mbox{\boldmath$r$}}
{\mbox{\boldmath$v$}}\phi_{i}/A$ are the mass and the average velocity of particle $i$, 
respectively. On the right hand side, ${\mbox{\boldmath$F$}}_i= \int d{\mbox{\boldmath$r$}}
\phi_i({\mbox{\boldmath$f$}_U+\mbox{\boldmath$f$}_T})$ is the force arising from the interparticle interaction and the torque, and 
${\mbox{\boldmath$K$}}_i=-\int d{\mbox{\boldmath$r$}}\phi_i \nabla\cdot {\stackrel{\leftrightarrow}{\mbox{\boldmath$\Pi$}}}
\cong -\int dS_{i}\hat{\mbox{\boldmath$n$}}_i\cdot {\stackrel{\leftrightarrow}{\mbox{\boldmath$\Pi$}}}$ 
the force exerted by the fluid. Here we use the following approximate relation  
$\int d{\mbox{\boldmath$r$}}\nabla\phi_i\cdot{\stackrel{\leftrightarrow}{\mbox{\boldmath$Q$}}}\cong -\int_{S_{i}}\hat{\mbox{\boldmath$n$}}_i 
dS_{i}\cdot{\stackrel{\leftrightarrow}{\mbox{\boldmath$Q$}}}$ 
for an arbitrary tensor ${\stackrel{\leftrightarrow}{\mbox{\boldmath$Q$}}}({\mbox{\boldmath$r$}})$, 
where $S_{i}$ is the surface of particle $i$ and $\hat{\mbox{\boldmath$n$}}_i$ is the unit outward normal vector to $S_i$. 
In practical numerical calculations, the on-lattice velocity field,  ${\mbox{\boldmath$v$}}({\mbox{\boldmath$r$}},t+\Delta t)$, 
is evaluated from the physical quantities at time $t$ by Eq. (\ref{Navier-Stokes}). 
Then we move particle $i$ off-lattice as a rigid body by ${\mbox{\boldmath$r$}}_i(t+\Delta t)={\mbox{\boldmath$r$}}_i(t)
+\Delta t{\mbox{\boldmath$V$}}_i(t+\Delta t)$, where $\Delta t$ is the time increment of the numerical integration. 

In our simulation, the units of length $\ell$ and time $\tau$ are related as $\tau=\ell^2/(\eta_{\rm \ell}/\rho)$, 
which sets both the scaled density and viscosity of the fluid region 
to unity. This $\tau$ is a time required for the fluid momentum to diffuse over a lattice size $\ell$.  
The units of stress and energy are $\bar\sigma=\rho(\ell/\tau)^2$ and $\bar\epsilon=\bar\sigma\ell^3$, 
respectively. Furthermore, we set $\eta_{\rm c}/\eta_{\rm \ell}=50$, $\Delta t=0.003$, $\ell=0.5 \xi=0.5$, and $a=6.4$. 
We confirmed that this choice yields reliable results by comparing them wit the analytical solution for a single particle rotation 
(see, e.g., Fig. \ref{fig:fig1}a and b).  
The simulation box used was typically $L^2=256^2$. To avoid cumbersome expressions,
we will use the same characters for the scaled variables below. We solve the equation of motion 
[Eq. (\ref{Navier-Stokes})] by the Marker-and-Cell (MAC) method with a staggered lattice 
under the periodic boundary condition. 

\noindent
{\bf Interparticle potentials.} 
To mimic rotating hard disks, we employ the Weeks-Chandler-Andersen (WCA) repulsive potential \cite{JDWeeksM}: 
$U_{jk}(r)=4 \epsilon \left\{(\sigma_{jk}/r)^{12}
-(\sigma_{jk}/r)^6
+1/4 \right\} {\rm for}\ \ r<2^{\frac{1}{6}}\sigma_{jk}$, otherwise 
$U_{jk}(r)=0$, where $\epsilon$ gives the energy scale, 
$\sigma_{jk}=(\sigma_{j}+\sigma_{k})/2$ 
and $\sigma_j$ represents the size of particle $j$.

\noindent
{\bf Characterization of structures.} 
The 2D radial distribution function $g(r)$ was calculated as 
\begin{eqnarray}
g(r)=\frac{1}{2\pi r\Delta r \rho (N-1)}\sum_{j \neq k} 
\delta(r-|\vec{r}_{jk}|), \label{eq:gr}
\end{eqnarray}
which is the ratio of the ensemble average of the number density of particles 
existing in the region $r \sim r+\Delta r$ to the average number density 
$\rho=N/L^2$. 
Here $N$ is the number of particles in the simulation box, 
whose side length is $L$, and $\Delta r$ is the increment of $r$. 

Similarly, the spatial correlation of $\Psi_6^{j}$ is calculated 
as \cite{NelsonB}
\begin{eqnarray}
g_6^{2D}(r)=\frac{L^2}{2\pi r\Delta rN(N-1)}\sum_{j \neq k}\delta(r-|\vec{r}_{jk}|)
\Psi_6^{j}\Psi_6^{k\ast}. \label{eq:g6}
\end{eqnarray}
The spatial correlation of the bond-orientational order 
can then be characterized by $g_6^{2D}(r)/g(r)$.

\vspace{\baselineskip}
\noindent
{\bf Acknowledgments}

\noindent
We are grateful to Akira Furukawa for discussion on the origin of hydrodynamic attractive interactions. 
We also thank Jun Russo for a critical reading of the manuscript. This study was partly
supported by Grants-in-Aid for Scientific Research (S) and Specially Promoted Research
from the Japan Society for the Promotion of Science (JSPS) and the Aihara Project, the
FIRST program from JSPS, initiated by the Council for Science and Technology Policy
(CSTP). 

\noindent
{\bf Author contributions} 

\noindent
H.T. proposed and supervised the study, Y.G. performed simulations, and H.T. wrote the manuscript.

\noindent
{\bf Competing financial interests:} The authors declare no competing financial interests. 

\clearpage

\noindent
{\bf SUPPLEMENTARY FIGURES}

\vspace{1cm}

\begin{center}
   \includegraphics[width =7cm]{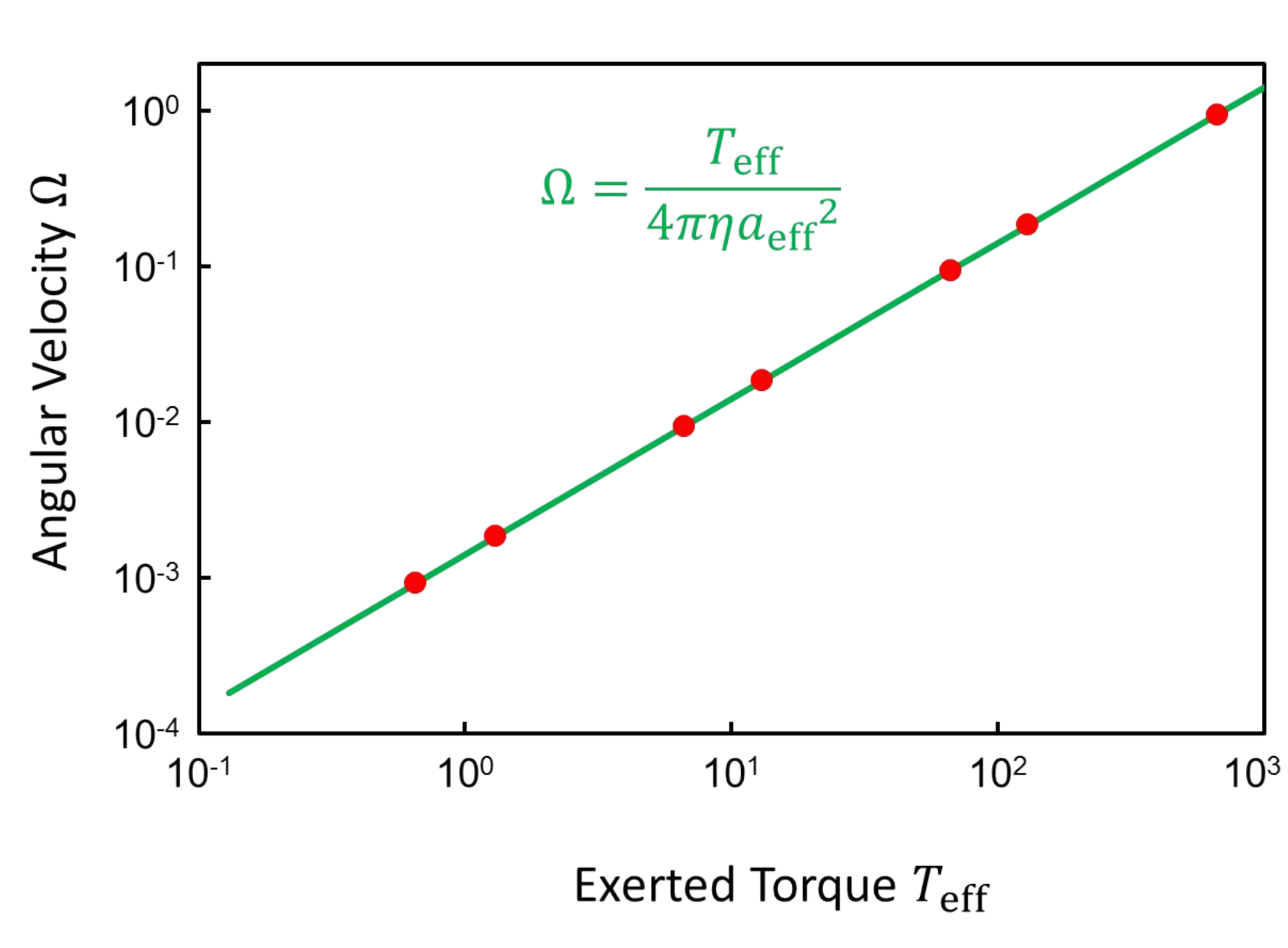}
  \end{center}
\noindent
{\bf Supplementary Figure 1. Relation between the exerted torque $T_{\rm eff}$ and the angular velocity $\Omega$ of a disk.} 

%  \label{fig:rfriction}

\vspace{1cm}

  \begin{center}
   \includegraphics[width =7cm]{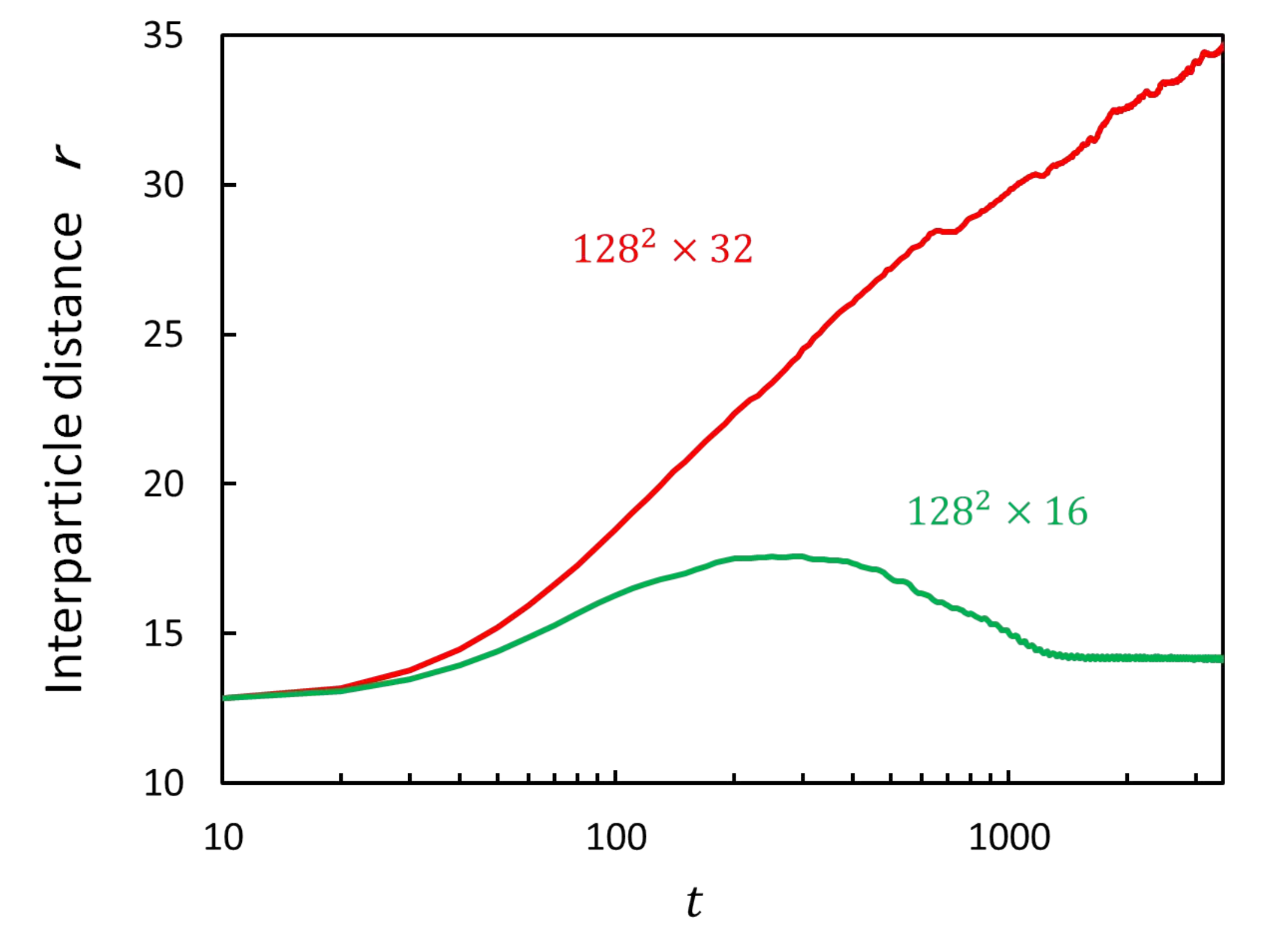}
  \end{center}
\noindent
{\bf Supplementary Figure 2. Switching of the effective hydrodynamic interaction between 3D and 2D.} 
  Decreasing the size of the third $z$ dimension of the quasi-2D simulation box ($128^2$ in $x$ and $y$ directions) from 32 to 16 
  leads to the switching of the interaction from repulsive to attractive, or the switching of the relaxation mode of vortex from cascade to anti-cascade mode. Accordingly, the two rotating disks 
  can have a stable rotating trajectory with a constant radius for the latter.

\vspace{1cm}
\noindent
{\bf SUPPLEMENTARY NOTES}

\vspace{1cm}

\noindent
{\bf Supplementary Note 1: The behaviour of a single rotating disc}

\noindent
As we show in the main text (Fig. 1a and b) our FPD method reasonably describe the basic behaviour of a rotating disk. 
Here we show an additional example which indicates the validity of our FPD method. 
Supplementary Figure 1 shows the relation between the effective exerted torque $T_{\rm eff}$ and the angular velocity $\Omega$ of a disk. 
We can see a perfect linear relation between them: $\Omega=T_{\rm eff}/(4 \pi \eta a_{\rm eff}^2)$. 
Here $T_{\rm eff}$ is the effective torque for our disk, which is described by the smooth profile function $\phi(\mathbf{r})$ and 
given by $T_{\rm eff}=T \int |\mathbf{r}|^2 \phi(\mathbf{r})^2 d\mathbf{r}/ \int |\mathbf{r}|^2 \phi(\mathbf{r}) d\mathbf{r}$, where $T$ is the 
exerted torque. From this analysis, we obtain $a_{\rm eff}=1.08 a$.

\vspace{1cm}
\noindent
{\bf Supplementary Note 2: The behaviour of a pair of rotating discs}

\noindent
As briefly discussed in the main text, a pair of rotating disks whose rotating axes are parallel have repulsive hydrodynamic interactions. 
So the distance between them monotonically increases with time (see the red curve in  Supplementary Fig. 2). 
However, decreasing the size of the third $z$ dimension of the quasi-2D simulation box ($128^2$ in $x$ and $y$ directions) from 32 to 16 
leads to the switching of the relaxation mode of vortex from cascade to anti-cascade mode. Accordingly, the two rotating disks 
can have a stable rotating trajectory with a constant radius (see the green curve in Supplementary Fig. 2). 
This is a consequence of the force balance between the repulsive Magnus force and the attractive hydrodynamic force. 
For 2D, thus, a pair of particles can form a stable rotating trajectory with a constant radius, as shown in Fig. 1d.  


\begin{thebibliography}{100}
\expandafter\ifx\csname url\endcsname\relax
  \def\url#1{\texttt{#1}}\fi
\expandafter\ifx\csname urlprefix\endcsname\relax\def\urlprefix{URL }\fi
\providecommand{\bibinfo}[2]{#2}
\providecommand{\eprint}[2][]{\url{#2}}

\bibitem{whitesides2002self}
\bibinfo{author}{Whitesides, G.~M.} \& \bibinfo{author}{Grzybowski, B.}
\newblock \bibinfo{title}{Self-assembly at all scales}.
\newblock \emph{\bibinfo{journal}{Science}} \textbf{\bibinfo{volume}{295}},
  \bibinfo{pages}{2418--2421} (\bibinfo{year}{2002}).

\bibitem{grzybowski2009self}
\bibinfo{author}{Grzybowski, B.~A.}, \bibinfo{author}{Wilmer, C.~E.},
  \bibinfo{author}{Kim, J.}, \bibinfo{author}{Browne, K.~P.} \&
  \bibinfo{author}{Bishop, K. J.~M.}
\newblock \bibinfo{title}{Self-assembly: from crystals to cells}.
\newblock \emph{\bibinfo{journal}{Soft Matter}} \textbf{\bibinfo{volume}{5}},
  \bibinfo{pages}{1110--1128} (\bibinfo{year}{2009}).

\bibitem{Marchetti2013}
\bibinfo{author}{Marchetti, M.~C.} \emph{et~al.}
\newblock \bibinfo{title}{Hydrodynamics of soft active matter}.
\newblock \emph{\bibinfo{journal}{Rev. Mod. Phys.}}
  \textbf{\bibinfo{volume}{85}}, \bibinfo{pages}{1143--1189}
  (\bibinfo{year}{2013}).

\bibitem{grzybowski2000dynamic}
\bibinfo{author}{Grzybowski, B.~A.}, \bibinfo{author}{Stone, H.~A.} \&
  \bibinfo{author}{Whitesides, G.~M.}
\newblock \bibinfo{title}{Dynamic self-assembly of magnetized, millimetre-sized
  objects rotating at a liquid--air interface}.
\newblock \emph{\bibinfo{journal}{Nature}} \textbf{\bibinfo{volume}{405}},
  \bibinfo{pages}{1033--1036} (\bibinfo{year}{2000}).

\bibitem{ishikawa2006interaction}
\bibinfo{author}{Ishikawa, T.} \& \bibinfo{author}{Hota, M.}
\newblock \bibinfo{title}{Interaction of two swimming Paramecia}.
\newblock \emph{\bibinfo{journal}{J. Exp. Biol.}}
  \textbf{\bibinfo{volume}{209}}, \bibinfo{pages}{4452--4463}
  (\bibinfo{year}{2006}).

\bibitem{lauga2009hydrodynamics}
\bibinfo{author}{Lauga, E.} \& \bibinfo{author}{Powers, T.~R.}
\newblock \bibinfo{title}{The hydrodynamics of swimming microorganisms}.
\newblock \emph{\bibinfo{journal}{Rep. Prog. Phys.}}
  \textbf{\bibinfo{volume}{72}}, \bibinfo{pages}{096601}
  (\bibinfo{year}{2009}).

\bibitem{baskaran2009statistical}
\bibinfo{author}{Baskaran, A.} \& \bibinfo{author}{Marchetti, M.~C.}
\newblock \bibinfo{title}{Statistical mechanics and hydrodynamics of bacterial
  suspensions}.
\newblock \emph{\bibinfo{journal}{Proc. Natl. Acad. Sci. U.S.A.}}
  \textbf{\bibinfo{volume}{106}}, \bibinfo{pages}{15567--15572}
  (\bibinfo{year}{2009}).

\bibitem{drescher2009dancing}
\bibinfo{author}{Drescher, K.} \emph{et~al.}
\newblock \bibinfo{title}{Dancing Volvox: Hydrodynamic bound states of swimming
  algae}.
\newblock \emph{\bibinfo{journal}{Phys. Rev. Lett.}}
  \textbf{\bibinfo{volume}{102}}, \bibinfo{pages}{168101}
  (\bibinfo{year}{2009}).

\bibitem{ramaswamy2010mechanics}
\bibinfo{author}{Ramaswamy, S.}
\newblock \bibinfo{title}{The Mechanics and Statistics of Active Matter}.
\newblock \emph{\bibinfo{journal}{Ann. Rev. Condens. Matter Phys.}}
  \textbf{\bibinfo{volume}{1}}, \bibinfo{pages}{323--45}
  (\bibinfo{year}{2010}).

\bibitem{wensink2012meso}
\bibinfo{author}{Wensink, H.~H.} \emph{et~al.}
\newblock \bibinfo{title}{Meso-scale turbulence in living fluids}.
\newblock \emph{\bibinfo{journal}{Proc. Natl. Acad. Sci. U.S.A.}}
  \textbf{\bibinfo{volume}{109}}, \bibinfo{pages}{14308--14313}
  (\bibinfo{year}{2012}).

\bibitem{Stark2014}
\bibinfo{author}{Z\"ottl, A.} \& \bibinfo{author}{Stark, H.}
\newblock \bibinfo{title}{Hydrodynamics determines collective motion and phase
  behavior of active colloids in quasi-two-dimensional confinement}.
\newblock \emph{\bibinfo{journal}{Phys. Rev. Lett.}}
  \textbf{\bibinfo{volume}{112}}, \bibinfo{pages}{118101}
  (\bibinfo{year}{2014}).

\bibitem{bialke2012crystallization}
\bibinfo{author}{Bialk{\'e}, J.}, \bibinfo{author}{Speck, T.} \&
  \bibinfo{author}{L{\"o}wen, H.}
\newblock \bibinfo{title}{Crystallization in a dense suspension of
  self-propelled particles}.
\newblock \emph{\bibinfo{journal}{Phys. Rev. Lett.}}
  \textbf{\bibinfo{volume}{108}}, \bibinfo{pages}{168301}
  (\bibinfo{year}{2012}).

\bibitem{grzybowski2001dynamic}
\bibinfo{author}{Grzybowski, B.~A.}, \bibinfo{author}{Jiang, X.},
  \bibinfo{author}{Stone, H.~A.} \& \bibinfo{author}{Whitesides, G.~M.}
\newblock \bibinfo{title}{Dynamic, self-assembled aggregates of magnetized,
  millimeter-sized objects rotating at the liquid-air interface: Macroscopic,
  two-dimensional classical artificial atoms and molecules}.
\newblock \emph{\bibinfo{journal}{Phys. Rev. E}} \textbf{\bibinfo{volume}{64}},
  \bibinfo{pages}{011603} (\bibinfo{year}{2001}).

\bibitem{grzybowski2002directed}
\bibinfo{author}{Grzybowski, B.~A.} \& \bibinfo{author}{Whitesides, G.~M.}
\newblock \bibinfo{title}{Directed dynamic self-assembly of objects rotating on
  two parallel fluid interfaces}.
\newblock \emph{\bibinfo{journal}{J. Chem. Phys.}}
  \textbf{\bibinfo{volume}{116}}, \bibinfo{pages}{8571} (\bibinfo{year}{2002}).

\bibitem{friese1998optical}
\bibinfo{author}{Friese, M. E.~J.}, \bibinfo{author}{Nieminen, T.~A.},
  \bibinfo{author}{Heckenberg, N.~R.} \& \bibinfo{author}{Rubinsztein-Dunlop,
  H.}
\newblock \bibinfo{title}{Optical alignment and spinning of laser-trapped
  microscopic particles}.
\newblock \emph{\bibinfo{journal}{Nature}} \textbf{\bibinfo{volume}{394}},
  \bibinfo{pages}{348--350} (\bibinfo{year}{1998}).

\bibitem{riedel2005self}
\bibinfo{author}{Riedel, I.~H.}, \bibinfo{author}{Kruse, K.} \&
  \bibinfo{author}{Howard, J.}
\newblock \bibinfo{title}{A self-organized vortex array of hydrodynamically
  entrained sperm cells}.
\newblock \emph{\bibinfo{journal}{Science}} \textbf{\bibinfo{volume}{309}},
  \bibinfo{pages}{300--303} (\bibinfo{year}{2005}).

\bibitem{schwarz2012phase}
\bibinfo{author}{Schwarz-Linek, J.} \emph{et~al.}
\newblock \bibinfo{title}{Phase separation and rotor self-assembly in active
  particle suspensions}.
\newblock \emph{\bibinfo{journal}{Proc. Natl. Acad. Sci. U.S.A.}}
  \textbf{\bibinfo{volume}{109}}, \bibinfo{pages}{4052--4057}
  (\bibinfo{year}{2012}).

\bibitem{lenz2003membranes}
\bibinfo{author}{Lenz, P.}, \bibinfo{author}{Joanny, J.~F.},
  \bibinfo{author}{J{\"u}licher, F.} \& \bibinfo{author}{Prost, J.}
\newblock \bibinfo{title}{Membranes with rotating motors}.
\newblock \emph{\bibinfo{journal}{Phys. Rev. Lett.}}
  \textbf{\bibinfo{volume}{91}}, \bibinfo{pages}{108104}
  (\bibinfo{year}{2003}).

\bibitem{gehrig2006nonlinear}
\bibinfo{author}{Gehrig, E.} \& \bibinfo{author}{Hess, O.}
\newblock \bibinfo{title}{Nonlinear dynamics and self-organization of rotary
  molecular motor ensembles}.
\newblock \emph{\bibinfo{journal}{Phys. Rev. E}} \textbf{\bibinfo{volume}{73}},
  \bibinfo{pages}{051916} (\bibinfo{year}{2006}).

\bibitem{llopis2008hydrodynamic}
\bibinfo{author}{Llopis, I.} \& \bibinfo{author}{Pagonabarraga, I.}
\newblock \bibinfo{title}{Hydrodynamic regimes of active rotators at fluid
  interfaces}.
\newblock \emph{\bibinfo{journal}{Eur. Phys. J. E}}
  \textbf{\bibinfo{volume}{26}}, \bibinfo{pages}{103--113}
  (\bibinfo{year}{2008}).

\bibitem{leoni2010dynamics}
\bibinfo{author}{Leoni, M.} \& \bibinfo{author}{Liverpool, T.~B.}
\newblock \bibinfo{title}{Dynamics and interactions of active rotors}.
\newblock \emph{\bibinfo{journal}{Europhys. Lett.}}
  \textbf{\bibinfo{volume}{92}}, \bibinfo{pages}{64004} (\bibinfo{year}{2010}).

\bibitem{goetze2010flow}
\bibinfo{author}{G{\"o}tze, I.~O.} \& \bibinfo{author}{Gompper, G.}
\newblock \bibinfo{title}{Flow generation by rotating colloids in planar
  microchannels}.
\newblock \emph{\bibinfo{journal}{Europhys. Lett.}}
  \textbf{\bibinfo{volume}{92}}, \bibinfo{pages}{64003} (\bibinfo{year}{2010}).

\bibitem{gotze2011dynamic}
\bibinfo{author}{G{\"o}tze, I.~O.} \& \bibinfo{author}{Gompper, G.}
\newblock \bibinfo{title}{Dynamic self-assembly and directed flow of rotating
  colloids in microchannels}.
\newblock \emph{\bibinfo{journal}{Phys. Rev. E}} \textbf{\bibinfo{volume}{84}},
  \bibinfo{pages}{031404} (\bibinfo{year}{2011}).

\bibitem{Gompper2014}
\bibinfo{author}{Yang, Y.}, \bibinfo{author}{Qiu, F.} \&
  \bibinfo{author}{Gompper, G.}
\newblock \bibinfo{title}{Self-organized vortices of circling self-propelled
  particles and curved active flagella}.
\newblock \emph{\bibinfo{journal}{Phys. Rev. E}} \textbf{\bibinfo{volume}{89}},
  \bibinfo{pages}{012720} (\bibinfo{year}{2014}).

\bibitem{fily2012cooperative}
\bibinfo{author}{Fily, Y.}, \bibinfo{author}{Baskaran, A.} \&
  \bibinfo{author}{Marchetti, M.~C.}
\newblock \bibinfo{title}{Cooperative self-propulsion of active and passive
  rotors}.
\newblock \emph{\bibinfo{journal}{Soft Matter}} \textbf{\bibinfo{volume}{8}},
  \bibinfo{pages}{3002--3009} (\bibinfo{year}{2012}).

\bibitem{eyink2006onsager}
\bibinfo{author}{Eyink, G.~L.} \& \bibinfo{author}{Sreenivasan, K.~R.}
\newblock \bibinfo{title}{Onsager and the theory of hydrodynamic turbulence}.
\newblock \emph{\bibinfo{journal}{Rev. Mod. Phys.}}
  \textbf{\bibinfo{volume}{78}}, \bibinfo{pages}{87} (\bibinfo{year}{2006}).

\bibitem{durkin2000experiments}
\bibinfo{author}{Durkin, D.} \& \bibinfo{author}{Fajans, J.}
\newblock \bibinfo{title}{Experiments on two-dimensional vortex patterns}.
\newblock \emph{\bibinfo{journal}{Phys. Fluid}} \textbf{\bibinfo{volume}{12}},
  \bibinfo{pages}{289} (\bibinfo{year}{2000}).

\bibitem{aref2003vortex}
\bibinfo{author}{Aref, H.}, \bibinfo{author}{Newton, P.~K.},
  \bibinfo{author}{Stremler, M.~A.}, \bibinfo{author}{Tokieda, T.} \&
  \bibinfo{author}{Vainchtein, D.~L.}
\newblock \bibinfo{title}{Vortex crystals}.
\newblock \emph{\bibinfo{journal}{Adv. Appl. Mech.}}
  \textbf{\bibinfo{volume}{39}}, \bibinfo{pages}{1--79} (\bibinfo{year}{2003}).

\bibitem{rasmussen2002dynamics}
\bibinfo{author}{Rasmussen, J.~J.}, \bibinfo{author}{Nielsen, A.~H.} \&
  \bibinfo{author}{Naulin, V.}
\newblock \bibinfo{title}{Dynamics of vortex interactions in two-dimensional
  flows}.
\newblock \emph{\bibinfo{journal}{Phys. Scr.}} \textbf{\bibinfo{volume}{2002}},
  \bibinfo{pages}{29} (\bibinfo{year}{2002}).

\bibitem{aref2007point}
\bibinfo{author}{Aref, H.}
\newblock \bibinfo{title}{Point vortex dynamics: A classical mathematics
  playground}.
\newblock \emph{\bibinfo{journal}{J. Math. Phys.}}
  \textbf{\bibinfo{volume}{48}}, \bibinfo{pages}{065401}
  (\bibinfo{year}{2007}).

\bibitem{FPD}
\bibinfo{author}{Tanaka, H.} \& \bibinfo{author}{Araki, T.}
\newblock \bibinfo{title}{Simulation method of colloidal suspensions with
  hydrodynamic interactions: Fluid particle dynamics}.
\newblock \emph{\bibinfo{journal}{Phys. Rev. Lett.}}
  \textbf{\bibinfo{volume}{85}}, \bibinfo{pages}{1338--1341}
  (\bibinfo{year}{2000}).

\bibitem{NelsonB}
\bibinfo{author}{Nelson, D.~R.}
\newblock \emph{\bibinfo{title}{Defects and Geometry in Condensed Matter
  Physics}} (\bibinfo{publisher}{Cambridge University Press., Cambridge},
  \bibinfo{year}{2002}).

\bibitem{montgomery2002experimental}
\bibinfo{author}{Montgomery, M.~T.}, \bibinfo{author}{Vladimirov, V.~A.} \&
  \bibinfo{author}{Denissenko, P.~V.}
\newblock \bibinfo{title}{An experimental study on hurricane mesovortices}.
\newblock \emph{\bibinfo{journal}{J. Fluid Mech.}}
  \textbf{\bibinfo{volume}{471}}, \bibinfo{pages}{1--32}
  (\bibinfo{year}{2002}).

\bibitem{le2002viscous}
\bibinfo{author}{Le~Dizes, S.} \& \bibinfo{author}{Verga, A.}
\newblock \bibinfo{title}{Viscous interactions of two co-rotating vortices
  before merging}.
\newblock \emph{\bibinfo{journal}{J. Fluid Mech.}}
  \textbf{\bibinfo{volume}{467}}, \bibinfo{pages}{389--410}
  (\bibinfo{year}{2002}).

\bibitem{yeo2007dynamic}
\bibinfo{author}{Yeo, K.}, \bibinfo{author}{Maxey, M.~R.} \&
  \bibinfo{author}{Karniadakis, G.~E.}
\newblock \bibinfo{title}{Dynamic self-assembly of spinning particles}.
\newblock \emph{\bibinfo{journal}{J. Fluid. Eng.}}
  \textbf{\bibinfo{volume}{129}}, \bibinfo{pages}{379} (\bibinfo{year}{2007}).

\bibitem{landau1987fluid}
\bibinfo{author}{Landau, L.~D.} \& \bibinfo{author}{Lifshitz, E.~M.}
\newblock \bibinfo{title}{Fluid Mechanics, Vol. 6}.
\newblock \emph{\bibinfo{journal}{Course of Theoretical Physics}}
  \bibinfo{pages}{227--229} (\bibinfo{year}{1987}).

\bibitem{tanaka1989digital}
\bibinfo{author}{Tanaka, H.}, \bibinfo{author}{Hayashi, T.} \&
  \bibinfo{author}{Nishi, T.}
\newblock \bibinfo{title}{Digital image analysis of droplet patterns in polymer
  systems: Point pattern}.
\newblock \emph{\bibinfo{journal}{J. Appl. Phys.}}
  \textbf{\bibinfo{volume}{65}}, \bibinfo{pages}{4480--4495}
  (\bibinfo{year}{1989}).

\bibitem{KAT}
\bibinfo{author}{Kawasaki, T.}, \bibinfo{author}{Araki, T.} \&
  \bibinfo{author}{Tanaka, H.}
\newblock \bibinfo{title}{Correlation between dynamic heterogeneity and
  medium-range order in two-dimensional glass-forming liquids}.
\newblock \emph{\bibinfo{journal}{Phys. Rev. Lett.}}
  \textbf{\bibinfo{volume}{99}}, \bibinfo{pages}{215701}
  (\bibinfo{year}{2007}).

\bibitem{krauth}
\bibinfo{author}{Bernard, E.~P.} \& \bibinfo{author}{Krauth, W.}
\newblock \bibinfo{title}{Two-step melting in two dimensions: First-order
  liquid-hexatic transition}.
\newblock \emph{\bibinfo{journal}{Phys. Rev. Lett.}}
  \textbf{\bibinfo{volume}{107}}, \bibinfo{pages}{155704}
  (\bibinfo{year}{2011}).

\bibitem{nguyen2013emergent}
\bibinfo{author}{Nguyen, N. H.~P.}, \bibinfo{author}{Klotsa, D.},
  \bibinfo{author}{Engel, M.} \& \bibinfo{author}{Glotzer, S.~C.}
\newblock \bibinfo{title}{Emergent collective phenomena in a mixture of hard
  shapes through active rotation}.
\newblock \emph{\bibinfo{journal}{Phys. Rev. Lett.}}
  \textbf{\bibinfo{volume}{112}}, \bibinfo{pages}{075701}
  (\bibinfo{year}{2014}).

\bibitem{voituriez2006generic}
\bibinfo{author}{Voituriez, R.}, \bibinfo{author}{Joanny, J.~F.} \&
  \bibinfo{author}{Prost, J.}
\newblock \bibinfo{title}{Generic phase diagram of active polar films}.
\newblock \emph{\bibinfo{journal}{Phys. Rev. Lett.}}
  \textbf{\bibinfo{volume}{96}}, \bibinfo{pages}{028102}
  (\bibinfo{year}{2006}).

\bibitem{terray2002microfluidic}
\bibinfo{author}{Terray, A.}, \bibinfo{author}{Oakey, J.} \&
  \bibinfo{author}{Marr, D. W.~M.}
\newblock \bibinfo{title}{Microfluidic control using colloidal devices}.
\newblock \emph{\bibinfo{journal}{Science}} \textbf{\bibinfo{volume}{296}},
  \bibinfo{pages}{1841--1844} (\bibinfo{year}{2002}).

\bibitem{bleil2006field}
\bibinfo{author}{Bleil, S.}, \bibinfo{author}{Marr, D. W.~M.} \&
  \bibinfo{author}{Bechinger, C.}
\newblock \bibinfo{title}{Field-mediated self-assembly and actuation of highly
  parallel microfluidic devices}.
\newblock \emph{\bibinfo{journal}{Appl. Phys. Lett.}}
  \textbf{\bibinfo{volume}{88}}, \bibinfo{pages}{263515}
  (\bibinfo{year}{2006}).

\bibitem{furukawa2010key}
\bibinfo{author}{Furukawa, A.} \& \bibinfo{author}{Tanaka, H.}
\newblock \bibinfo{title}{Key role of hydrodynamic interactions in colloidal
  gelation}.
\newblock \emph{\bibinfo{journal}{Phys. Rev. Lett.}}
  \textbf{\bibinfo{volume}{104}}, \bibinfo{pages}{245702}
  (\bibinfo{year}{2010}).

\bibitem{JDWeeksM}
\bibinfo{author}{Weeks, J.~D.}, \bibinfo{author}{Chandler, D.} \&
  \bibinfo{author}{Andersen, H.~C.}
\newblock \bibinfo{title}{Role of repulsive forces in determining the
  equilibrium structure of simple Llquids}.
\newblock \emph{\bibinfo{journal}{J. Chem. Phys.}}
  \textbf{\bibinfo{volume}{54}}, \bibinfo{pages}{5237--5247}
  (\bibinfo{year}{1971}).

\end{thebibliography}
\end{document}